\documentclass[prd,aps,twocolumn,a4paper,floatfix,nofootinbib]{revtex4-1}

\usepackage[utf8]{inputenc}
\usepackage{graphicx,psfrag}
\usepackage{mathrsfs}
\usepackage{amsmath,amsfonts,amssymb}
\usepackage{multirow}
\usepackage{comment}
\usepackage{xcolor}
\usepackage{hyperref}
\hypersetup{
    colorlinks = true,
    linkcolor = {blue},
    citecolor = {blue},
    urlcolor = {blue},
    linkbordercolor = {white},
    }

\newcommand{\be}{\begin{equation}}
\newcommand{\ee}{\end{equation}}
\newcommand{\bea}{\begin{eqnarray}}
\newcommand{\eea}{\end{eqnarray}}
\newcommand{\bel}{\begin{align}}
\newcommand{\eel}{\end{align}}

\def\l{\ell}
\def\lm{\ell m}

\def\GMc2{{\rm G M_{\odot} c^{-2}}}

\def\cS{\mathcal{S}}
\def\cT{\mathcal{T}}

\usepackage{color}

\definecolor{cyan}{rgb}{0,0.9,0.9}
\definecolor{orange}{rgb}{0.9,0.5,0}
\definecolor{magenta}{rgb}{1,0,1}
\definecolor{purple}{rgb}{0.8,0.4,0.8}
\definecolor{gray}{rgb}{0.8242,0.8242,0.8242}
\definecolor{mgreen}{rgb}{0.1,0.8,0.1}

\DeclareMathOperator{\PSD}{PSD}

\usepackage[normalem]{ulem}

\begin{document}

\title{Gravitational waves and mass ejecta from binary neutron star mergers:
Effect of large eccentricities}

\author{Swami Vivekanandji \surname{Chaurasia}$^{1}$}
\author{Tim \surname{Dietrich}$^{2}$}
\author{Nathan K. \surname{Johnson-McDaniel}$^3$}
\author{Maximiliano \surname{Ujevic}$^4$}
\author{Wolfgang \surname{Tichy}$^5$}
\author{Bernd \surname{Br\"ugmann}$^1$}

\affiliation{${}^1$ Theoretical Physics Institute, University of Jena, 07743 Jena, Germany}  
\affiliation{${}^2$ Nikhef, Science Park 105, 1098 XG Amsterdam, The Netherlands}
\affiliation{${}^3$ DAMTP, Centre for Mathematical 
Sciences, Wilberforce Road, Cambridge, CB3 0WA,
United Kingdom}
\affiliation{${}^4$ Centro de Ci\^encias Naturais e Humanas, Universidade Federal do ABC, 09210-170, Santo Andr\'e, S\~ao Paulo, Brazil}
\affiliation{${}^5$ Department of Physics, Florida Atlantic University, Boca Raton, FL 33431 USA}

\date{\today}

\begin{abstract}
As current gravitational wave (GW) detectors increase in sensitivity, 
and particularly as new instruments
are being planned,
there is the possibility that ground-based GW detectors will observe GWs from
highly eccentric neutron star binaries.
We present the first detailed study of highly eccentric BNS systems 
with full (3+1)D numerical relativity simulations 
using consistent initial conditions, 
i.e., setups which are in agreement with the Einstein equations 
and with the equations of general relativistic hydrodynamics in equilibrium. 
Overall, our simulations cover two different equations of state (EOSs), two different spin configurations, 
and three to four different initial eccentricities for each pairing of EOS and spin.
We extract from the simulated waveforms the frequency of the $\textit{f}$-mode oscillations
induced during close encounters before the merger of the two stars.
The extracted frequency is in good agreement with $\textit{f}$-mode
oscillations of individual stars for the irrotational cases, which allows an independent
measure of the supranuclear equation of state not accessible
for binaries on quasicircular orbits. 
The energy stored in these $f$-mode oscillations can be as large as $10^{-3}M_\odot \sim 10^{51}$~erg,
even with a soft EOS. 
In order to estimate the stored energy, we also examine the effects of mode mixing
due to the stars' offset from the origin on the $f$-mode contribution to the GW signal.
While in general (eccentric) neutron star mergers produce bright electromagnetic counterparts,
we find that for the considered cases with fixed initial separation the luminosity decreases when the eccentricity becomes too large, due to 
a decrease of the ejecta mass.
Finally, the use of consistent initial configurations also allows us
to produce high-quality waveforms for different eccentricities which can be
used as a test bed for waveform model development of highly eccentric binary
neutron star systems.
\end{abstract}


\maketitle

\section{Introduction}
\label{sec:intro}
The detection of the binary black hole (BBH) merger
GW150914~\cite{abbott:2016blz} in 2015 has initiated the era of
gravitational wave (GW) astronomy. Since then, a number of additional BBH systems have
been
detected~\cite{abbott:2016nmj,TheLIGOScientific:2016pea,abbott:2017vtc,abbott:2017oio,abbott:2017gyy}. 
Apart from the detections of BBHs,
the spectacular observation of
both GWs and electromagnetic (EM) radiation from
a binary neutron star (BNS) merger in August 2017,
GW170817~\cite{TheLIGOScientific:2017qsa,GBM:2017lvd,
Monitor:2017mdv}, has been
a scientific breakthrough for multi-messenger astronomy.
With the planned upgrades of the advanced GW detectors LIGO and VIRGO~\cite{LIGO,Virgo}
and  the upcoming KAGRA~\cite{KAGRA} detector in Japan, the near future
of GW astronomy is bright
and multiple detections of compact binaries
are expected in the coming
years~\cite{Aasi:2013wya,Abbott:2016nhf}. 
In addition to those improvements, the possibility of a 3rd generation (3G)
of GW detectors, such as the Einstein Telescope (ET)~\cite{Hild:2008ng,
Punturo:2010zz,Ballmer:2015mvn} and Cosmic Explorer~\cite{Evans:2016mbw},
is exciting,
since 3G detectors are expected to be ten times more sensitive than
currently operating detectors.
3G observatories would thus
not only provide a significantly larger number of detections,
but also give the possibility of detecting systems and signals not observable by current interferometers,
e.g., the postmerger phase of the remnant formed after the collision~\cite{Clark:2014wua,Clark:2015zxa,
Radice:2016gym,Radice:2016rys,Abbott:2017dke}
or highly eccentric BNS systems.

In fact, in anticipation of an increased number of BNS detections
with higher signal-to-noise ratio
and the possibility to detect such extreme
configurations as precessing BNSs, highly eccentric systems,
or high-mass ratio mergers, it becomes even more important
to model compact binary waveforms accurately
to allow for a precise measurement of the source parameters.
In order to extract information from a measured GW signal,
the data are compared with fast-to-evaluate waveform models, which need to cover the entire BNS parameter
space to allow for the accurate estimation of the binary parameters for all possible systems.
While there has been progress in improving waveform models for (moderately) eccentric BBH
systems~\cite{Hinderer:2017jcs,Cao:2017ndf,Hinder:2017sxy,Huerta:2017kez} and
very recently Ref.~\cite{Yang:2018bzx} proposed a series of studies
for waveform model development which will capture the dynamics of BNS systems
on eccentric orbits up to the merger of the stars, there is currently no waveform model 
for BNS systems on highly eccentric orbits. 
In numerical relativity (NR), most groups have focused on the simulations of
quasi-circular BNS. This restriction is reasonable since the vast majority of
systems are expected to have only a small eccentricity once the system enters the
LIGO band due to the decay of eccentricity by the emission of gravitational
radiation~\cite{Peters:1964zz,Kowalska:2010qg}. However, different
channels have been suggested for the formation of NS binaries that may retain
non-negligible eccentricity when they merge.

One such proposed channel considers the capture of two
initially mutually unbound neutron stars in a dense stellar system such as
a nuclear star cluster via the emission of gravitational radiation
during a close non-merging encounter~\cite{Tsang:2013mca}.
Reference~\cite{Tsang:2013mca} reports an estimate\footnote{Note that~\cite{Tsang:2013mca} mentions 
that the stated rates may be overestimated by a factor of $\sim 4$.} of the volume
rate of such encounters of 
$\sim 0.003$--$6$~Gpc$^{-3}$~yr$^{-1}$.
To obtain a possible detection rate, we need to incorporate the sensitive
volume of ET. We assume two different scenarios 
to estimate the sensitive volume, which should bracket the sensitive volumes
appropriate for realistic data analysis techniques (see, e.g.,~\cite{East:2012xq} for initial work on such
techniques). Specifically, we assume
(i) the use of an unmodelled search for the kHz GW radiation emitted during the merger, for which
one obtains a range of $\sim 20$~Mpc from Fig.~21 in~\cite{Abernathy:2011ET},\footnote{Note
that the radiated energy of $0.05 M_\odot$ used
to construct Fig.~21 in~\cite{Abernathy:2011ET} is compatible with the mergers
of binaries with a soft EOS we consider, 
but we have to reduce the distances by a factor 
of $\sim 3$, due to the longer times over which the energy 
is emitted, using the scaling in that paper's Eq.~(13).} and 
(ii) that it is possible to construct a template bank of sufficiently accurate waveforms
to enable a matched-filter search for these systems, giving a range of $\sim 8$~Gpc, where we estimate the range
for highly eccentric binaries to be about half that for quasicircular binaries (obtained from Fig.~18 in~\cite{Abernathy:2011ET}), since the matched
filtering range for Advanced LIGO for highly eccentric binaries shown in Fig.~16 of~\cite{East:2012xq} is (on the larger side)
about half that for quasicircular binaries quoted in~\cite{Aasi:2013wya}. 

This leads to detection rates from this channel of the order of 
$10^{-7}$--$10^{-4}$~yr$^{-1}$ for burst searches [type (i)] and 
$10^{-1}$--$10^3$~yr$^{-1}$ for matched filter searches [type (ii)], where we computed the comoving volume within the range
using~\cite{Wright:2006up}.

Another dominant channel for eccentric BNS inspirals is suggested in
\cite{Samsing:2013kua}, who consider binary-single star interactions in globular
clusters (GCs) and find the rates to be $\sim 0.7$~Gpc$^{-3}$~yr$^{-1}$ for
typical GCs containing $\sim 10^3$ NSs and assuming that 30\% of these are in
binaries. This would give eccentric BNS merger rates observable by ET on the
order of $10^{-5}$ and $10^2$~yr$^{-1}$ for type (i) and (ii) searches, respectively.
These numbers show that it might be possible to detect a few highly eccentric BNS mergers per
year with an operating ET. We note that even if one is not able to perform a matched filter search,
the unmodeled search numbers are surely pessimistic, as one should be able to do better than a
purely unmodeled search for just the postmerger signal. Additionally, development of an eccentric BNS waveform model
allowing for matched filtered searches would significantly boost the possibility 
of detecting highly eccentric BNS systems.\footnote{Another 
possible channel stated in~\cite{Chirenti:2016xys} for cases
where one is interested in studying tidal mode excitations of the NS
matter are close periastron passages in nuclear star clusters. They find $\sim 10^{-3} - 1$ per year within the ET
sensitive volume.}\\ 

As eccentric BNSs merge, they may produce brighter EM emission
than quasicircular mergers, making it important to understand such systems from a
multimessenger perspective.
Furthermore, the gravitational waveforms for highly eccentric binaries significantly differ
from the classical chirp signal of quasicircular inspirals with its slowly
increasing amplitude and frequency.
On highly eccentric orbits, each encounter of the stars
leads to a burst of GW radiation, first studied using Newtonian orbits together
with leading order relativistic expressions for radiation and the evolution of orbital 
parameters~\cite{Peters:1963ux,Turner:1977a}. Over the last decade, there have been numerical
simulations of highly eccentric BBH~\cite{Pretorius:2007jn,Sperhake:2007gu,Gold:2009hr,Gold:2012tk} and
BHNS~\cite{Stephens:2011as,East:2011xa,East:2015yea} systems in
full general relativity (GR).
There have been studies exploring highly eccentric (also known as dynamical
capture) BNS in Newtonian gravity~\cite{Lee:2009ca,Rosswog:2012wb}
and in full GR simulations
with approximate initial data (ID) using a simple superposition of two boosted
NSs~\cite{Gold:2011df,Radice:2016dwd,Papenfort:2018bjk}, and in \cite{East:2012ww,East:2015vix,
Paschalidis:2015mla,East:2016zvv}
with constraint solved ID, but not in hydrodynamical equilibrium.

Results for BHNS/BNS systems indicate a strong variability in properties of the wave and
matter dynamics as a function of the eccentricity (or impact parameter).
For example, for the eccentric BHNS systems studied in~\cite{Stephens:2011as}, the remnant
disk masses range from nearly zero up to $\sim 0.3~M_{\odot}$, the unbound masses vary from zero
to $\sim 0.15~M_{\odot}$ (computed from Table~1 of~\cite{Stephens:2011as} using the baryonic mass
of $1.49 M_\odot$ from Table~II of~\cite{Read:2009yp}). The energy and angular momentum emitted during the
nonmerging first encounters are also found to vary by an order of magnitude depending on the impact
parameter. The dynamical capture BNSs studied in~\cite{Radice:2016dwd} (the models labeled `HY\_RPX' in Table~1)
also show variability in the type of merger remnant with impact parameter. Additionally, the unbound masses range from
$\sim0.0004$--$0.125~M_{\odot}$ as the impact parameter is varied, with corresponding variations in the signature of
the electromagnetic counterparts.

An important result concerning BNS (or BHNS) is that eccentricity leads to tidal
interactions that can excite oscillations of the stars, which in turn generate their own
characteristic GW signal \cite{Turner:1977b,Kokkotas:1995xe,Chirenti:2016xys,Parisi:2017kgx,Yang:2018bzx}.

Neither the orbital nor stellar GWs have been studied so far
for eccentric BNS orbits in GR with \textit{consistent} ID, i.e.,
data which fulfill the Einstein constraint equations and the
equations of general relativistic hydrodynamics in equilibrium,
except for initial explorations with a simple equation of state in~\cite{Moldenhauer:2014yaa},
which still used an approximation for the velocity of the fluid, as opposed to solving for the
velocity potential. The problems with using inconsistent ID are constraint violations,
in cases for which the Einstein constraints are not fulfilled for the ID, or
spurious matter density oscillations if the fluid is not in equilibrium.
Those oscillations potentially spoil the quantitative analysis of matter
oscillations which arise due to the encounters in eccentric BNSs.

Currently, there are a number of advanced solvers computing BNS initial data
that are capable of exploring certain portions of the parameter space,
e.g., the publicly available spectral code 
LORENE~\cite{Bonazzola:1998yq,Gourgoulhon:2000nn,LoreneCode},
the Princeton group's multigrid solver~\cite{East:2012zn},
BAM's multigrid solver~\cite{Moldenhauer:2014yaa},
the COCAL code~\cite{Tsokaros:2015fea},
SpEC's pseudospectral solver Spells~\cite{Foucart:2008qt,Tacik:2015tja},
and the pseudospectral code SGRID~\cite{Tichy:2009yr,Tichy:2011gw,Tichy:2012rp,Dietrich:2015pxa}.
Unfortunately, all of these solvers are incapable of reaching certain portions
of the possible BNS parameter space. In particular, most of them
cannot generate consistent initial data with specified high eccentricities.
Only SGRID, the private LORENE version of Ref.~\cite{Kyutoku:2014yba}, and the Spells code
(cf.~\cite{Tacik:2015tja}) allow for adjusted orbital eccentricities for eccentric BNS 
simulations in a framework where the fluid equations are solved consistently.\footnote{However, note that so far this control is only
used to reduce eccentricity in Refs.~\cite{Kyutoku:2014yba,Tacik:2015tja}}.

In this paper we present the first detailed study of highly eccentric BNS systems using consistent ID
extending the results of~\cite{Dietrich:2016hky,Dietrich:2016lyp}
in which we have already studied the effects of the mass ratio and spin.
We consider equal-mass binaries at fixed initial separation
($\sim 165$~km) with two different EOSs
and we vary the initial eccentricity. 
The chosen EOSs, SLy and MS1b, are reasonably representative of soft and stiff
EOSs, respectively. Although analysis of the GW signal GW170817
~\cite{Abbott:2018wiz,Abbott:2018exr,TheLIGOScientific:2017qsa}
shows that MS1b predicts tidal deformabilities that are too large to agree with
observation (since it gives stars that are not very compact), using this EOS
allows us to understand the behavior
of less compact stars.

The article is structured as follows: In Sec.~\ref{sec:simu},
we give a short description of the numerical methods and describe important
quantities used to analyze our simulations. Section~\ref{sec:config} summarizes
the properties of the binaries we simulate.
Section~\ref{sec:dyn} deals with the dynamics of the simulations,
where in particular we focus on the conservative dynamics of the BNS
system and the merger remnant.
We discuss the properties of the dynamical ejecta and EM counterparts in Sec.~\ref{sec:ejectaandEM}.
In Sec.~\ref{sec:GWs}, we investigate the properties of the GW signal using spectrograms, and
also consider the NS $f$-mode oscillations (including the effects of mode mixing due to the stars'
displacement from the origin and an estimate of the energy stored in these oscillations) and the postmerger GW frequencies. We conclude in Sec.~\ref{sec:summary}.
In Appendix~\ref{app:conv} we test the accuracy of our simulations with respect to
conserved quantities, the constraints and the waveforms.

Throughout this work we use geometric units,
setting $c = G = M_{\odot} = 1$, but occasionally give quantities in 
astrophysical units to allow for easier interpretation.
Spatial indices are denoted by Latin letters running from 1 to 3 and Greek
letters are used for spacetime indices running from 0 to 3.

All but two simulations we present here are already publicly available 
in the CoRe database of binary neutron star merger 
waveforms~\cite{Dietrich:2018phi,CoRe}. The remaining waveforms will 
be made available in the near future.

\section{Methods}
\label{sec:simu}

\subsection{Initial configurations}
\label{ssec:initial}

Our initial configurations are constructed with the pseudospectral SGRID
code~\cite{Tichy:2006qn,Tichy:2009yr,Tichy:2009zr,Dietrich:2015pxa}.
SGRID uses the conformal thin sandwich formalism~\cite{Wilson:1995uh,Wilson:1996ty,York:1998hy}
in combination with the constant rotational velocity
approach~\cite{Tichy:2011gw,Tichy:2012rp} to construct
quasiequilibrium configurations of spinning neutron stars and
the methods presented in~\cite{Moldenhauer:2014yaa,Dietrich:2015pxa}
to allow for eccentric BNSs\footnote{This method does not include
star excitations in the initial data from any previous encounters.
However, any such initial oscillations will be smaller by at least
an order of magnitude compared to those later in the evolution, so the
absence of initial oscillations
does not affect the later oscillations very significantly.}.
In particular we assume for each star in the eccentric BNS the approximate \textit{helliptical}
Killing vectors (where the name denotes a combination of helical and elliptical motion):
\begin{subequations}
\label{eq:KV}
\begin{equation}
k_{1,2}^{\alpha} = t^\alpha + \Omega [(x-x_{c_{1,2}}) y^\alpha - y x^\alpha] 
+ \frac{v_r}{r_{12}} r^\alpha 
\end{equation}
with
\begin{equation}
 x_{c_{1,2}} = x_{\rm CM} + e (x_{1,2}-x_{\rm CM})
\end{equation}
\end{subequations}
the positions of the centers of the inscribed circles approximating the stars' orbits.
Here the stars (labeled by $1$ and $2$) start on the $x$-axis with positions $x_{1,2}$,
$\mathbf{t} = \partial_t$, $\mathbf{x}= \partial_x$,
$\mathbf{y} = \partial_y$ refer to the Cartesian coordinate vectors,
$r^\alpha= (0,x,y,z)$ and $r_{12}=\left\lvert x_1 - x_2 \right\rvert$ is 
the distance between the star centers.
The parameters $e$ and $v_r$ define the eccentricity and radial velocity, respectively---we
take $v_r = 0$ here. Additionally, $x_\text{CM}$ and $\Omega$ denote the system's center-of-mass and
angular velocity parameter, which are both determined by the force-balance equation, as discussed
in~\cite{Moldenhauer:2014yaa,Dietrich:2015pxa}, which give a more detailed discussion about the
construction of eccentric BNS configurations.

SGRID's computational domain is divided into six patches
(Fig.~1 of \cite{Dietrich:2015pxa}) including spatial infinity, which
allows us to impose exact boundary conditions.
We employ $n_A = n_B=28$, $n_\varphi=8$, $n_{\rm Cart} = 24$ points for
the spectral grid and give details of the configurations studied in Table~\ref{tab:config}.

\subsection{Evolutions}
Numerical relativity simulations are performed with the BAM
code~\cite{Brugmann:2008zz,Thierfelder:2011yi,Dietrich:2015iva,
Bernuzzi:2016pie}.
The Einstein equations are written in 3+1 form using the BSSNOK
evolution system~\cite{Nakamura:1987zz,Shibata:1995we,Baumgarte:1998te}.
The (1+log)-lapse and gamma-driver-shift conditions are employed for the
evolutions~\cite{Bona:1994a,Alcubierre:2002kk,vanMeter:2006vi}.
The general relativistic hydrodynamics (GRHD)
equations are solved in conservative form by
defining Eulerian conservative variables from the rest-mass density
$\rho$, pressure $p$, specific internal energy $\epsilon$, and 3-velocity, $v^i$.
The system is closed by an EOS for which we use piecewise-polytropic 
fits of the SLy and MS1b EOSs; see~\cite{Read:2009yp}.
We include thermal effects by adding an additional thermal pressure
of the form $p_{\rm th} = \rho \epsilon (\Gamma_\text{th}-1)$ with
$\Gamma_{\rm th}=1.75$, cf.~\cite{Bauswein:2010dn}.

The numerical domain is divided into a hierarchy of cell centered nested Cartesian grids.
The hierarchy consists of $L$ levels of refinement labeled by $l = 0$, $\cdots$, $L-1$.
Each refinement level $l$ has one or more Cartesian grids with constant grid spacing
$h_l$ and $n$ (or $n^{\rm mv}$) points per direction.
The refinement factor is two such that $h_l = h_0/2^l$.
The grids are properly nested, i.e., the coordinate extent of any grid at level
$l$, $l > 0$, is completely covered by the grids at level $l-1$.
Some of the mesh refinement levels $l > l^{\rm mv}$
can be dynamically moved and adapted during the time evolution
according to the technique of ``moving boxes''; for this work we set
$l^{\rm mv} = 5$.
The BAM grid setup considered in this work consists of nine refinement
levels.
Details about the different grid configurations employed in this work are given in Table~\ref{tab:grid};
the grid configurations are labeled R1, R2, R3, R4, ordered by increasing resolution.

Time integration is performed with the method of lines using explicit fourth 
order Runge-Kutta integrators.
Derivatives of metric fields
are approximated by fourth-order finite differences, while a high-resolution-shock-capturing 
scheme based on primitive reconstruction and the local Lax-Friedrichs (LLF) 
central scheme for the numerical fluxes is adopted for the 
matter~\cite{Thierfelder:2011yi}. Primitive reconstruction is performed 
with the fifth order WENOZ scheme of~\cite{Borges:2008a}.

\begin{table*}[t]
\caption{BNS configurations. 
The first column gives the configuration name and the second column gives the corresponding
CoRe database ID.
The next 5 columns provide the physical properties of the single stars:
the EOS, the gravitational masses of the individual stars $M^{A,B}$, the baryonic masses of
the individual stars $M _b ^{A,B}$, the stars' dimensionful and dimensionless spins $S^{A,B}$ and $\chi ^{A,B}$.
The last 5 columns give the input eccentricity $e$ [Eq.~\eqref{eq:KV}],
the 3PN eccentricity $\hat{e}_{\rm 3PN}$ [Eq.~(\ref{eq:3PNecc})], the initial GW
frequency $M\omega^0 _{22}$, the Arnowitt-Deser-Misner (ADM) mass $M_\text{ADM}$, and the ADM angular
momentum $J_\text{ADM}$. The configurations evolved with different resolutions
(cf.~Table~\ref{tab:grid}) are marked with asterisks (*).}
\label{tab:config} 
\centering
\begin{tabular}{c|c|ccccc|ccccc}
\toprule
Name & CoRe DB ID & EOS & $M^{A,B}$ & $M^{A,B}_b$ & $S^{A,B}$ & $\chi^{A,B}$ & $e$ & $\hat{e}_{\rm 3PN}$ & $M\omega ^0 _{22}$ & $M_\text{ADM}$ & $J_\text{ADM}$ \\ 
\hline 
SLy$-e0.40 ^{00}$ & BAM:0112 & SLy &  $1.357558$ & $1.504$  & $0.0000$  & $0.0000$ & $0.40$ & $0.52$ & $0.0106$ & $2.702935$ & $8.8086$ \\
SLy$-e0.45 ^{00*}$ & BAM:0113 & SLy & $1.357558$ & $1.504$ & $0.0000$ & $0.0000$ & $0.45$  & $0.55$ & $0.0110$ & $2.702408$ & $8.4514$ \\ 
SLy$-e0.50 ^{00*}$ & BAM:0114 & SLy & $1.357558$ & $1.504$ & $0.0000$ & $0.0000$ & $0.50$ & $0.60$ &  $0.0116$ & $2.701945$ & $8.0487$ \\
SLy$-e0.60 ^{00}$ & BAM:0115 & SLy & $1.357558$ & $1.504$ & $0.0000$ & $0.0000$ & $0.60$ & $0.69$ &  $0.0129$ & $2.700947$ & $7.1828$ \\
\toprule
SLy$-e0.40 ^{\uparrow\uparrow}$ & BAM:0116 & SLy & $1.358097$ & $1.504$ & $0.1767$ & $0.0958$ & $0.40$ & $0.50$ & $0.0107$ & $2.703786$ & $9.2346$ \\
SLy$-e0.45 ^{\uparrow\uparrow*}$ & BAM:0117 & SLy & $1.358097$ & $1.504$ & $0.1767$ & $0.0958$ & $0.45$ & $0.54$ & $0.0111$ & $2.703281$ & $8.8478$ \\ 
SLy$-e0.50 ^{\uparrow\uparrow}$ & BAM:0118 & SLy & $1.358097$ & $1.504$ & $0.1767$ & $0.0958$ & $0.50$ & $0.59$ & $0.0117$ & $2.702773$ & $8.4440$ \\
SLy$-e0.60 ^{\uparrow\uparrow}$ & BAM:0119 & SLy & $1.358097$ & $1.504$ & $0.1767$ & $0.0958$ & $0.60$ & $0.69$ & $0.0130$ & $2.701827$ & $7.5714$ \\
\toprule

MS1b$-e0.45 ^{00}$ & BAM:0074 & MS1b & $1.380825$ & $1.504$ & $0.0000$ & $0.0000$ & $0.45$ & $0.47$ & $0.0119$ & $2.748643$ & $9.1766$ \\ 
MS1b$-e0.50 ^{00}$  & BAM:0075 & MS1b & $1.380825$& $1.504$ & $0.0000$ & $0.0000$ & $0.50$ & $0.53$ & $0.0125$ & $2.748120$ & $8.7322$ \\
MS1b$-e0.60 ^{00}$ & BAM:0076 & MS1b & $1.380825$ & $1.504$ & $0.0000$ & $0.0000$ & $0.60$ & $0.64$ & $0.0139$ & $2.747100$ & $7.7936$\\
\toprule

MS1b$-e0.45 ^{\uparrow\uparrow}$ & BAM:0077 & MS1b & $1.381270$ & $1.504$ & $0.2016$ & $0.1056$ & $0.45$ & $0.48$ & $0.0119$ & $2.749629$ & $9.5789$ \\ 
MS1b$-e0.50 ^{\uparrow\uparrow}$ & BAM:0078 & MS1b & $1.381270$ & $1.504$ & $0.2016$ & $0.1056$ & $0.50$ & $0.53$ & $0.0125$ & $2.749104$ & $9.1487$ \\
MS1b$-e0.60 ^{\uparrow\uparrow}$ & BAM:0079 & MS1b & $1.381270$ & $1.504$ & $0.2016$ & $0.1056$ & $0.60$ & $0.64$ & $0.0139$ & $2.748047$ & $8.2188$ \\
\hline
\end{tabular} 
\end{table*}

\begin{table}[tp]
\caption{Grid configurations. 
The columns refer to: the resolution name, the number of points in the nonmoving boxes $n$,
the number of points in the moving boxes $n^\text{mv}$,
the grid spacing in the finest level $h_8$ covering the NS diameter,
the grid spacing in the coarsest level $h_0$, and
the outer boundary position $R_0$. The grid spacing and
the outer boundary position are given in units of $M_{\odot}$}
\label{tab:grid} 
\centering
\begin{tabular}{c|ccccc}
Name & $n$ & $n^\text{mv}$ & $h_8$ & $h_0$ & $R_0$ \\
\toprule
R1 & $128$ & $64$ & $0.219$ & $56.00$ & $3612.000$\\
\hline
R2 & $192$ & $96$ & $0.146$ & $37.38$ & $3606.784$\\
\hline
R3 & $256$ & $128$ & $0.110$ & $28.03$ & $3602.112$ \\
\hline
R4 & $320$ & $160$ & $0.088$ & $22.43$ & $3599.309$ \\
\end{tabular}
\end{table}

\subsection{Simulation analysis}
\label{ssec:sim_analysis}
Most of our analysis tools were summarized in 
Refs.~\cite{Dietrich:2016hky,Dietrich:2016lyp}, including
the computation of the ejecta properties, the disk masses,
merger remnant characterizations, and the extraction of GWs.\\

\paragraph*{\textbf{Ejecta computation.}}
For the analysis in this paper, we extended our ejecta computation,
which was previously purely based on the volume integration
of the unbound matter, $M_{\rm ej}^{\mathcal{V}}$.
Now, as an alternative method, we compute the matter flux of the 
unbound material across coordinate spheres sufficiently far
from the system,
$M_{\rm ej}^{\mathcal{S}}$; see
e.g.,~\cite{Kastaun:2014fna,Radice:2016dwd}.
In general, we mark matter as unbound if it fulfills
\begin{equation}
\label{eq:unbound}
u_t<-1 \quad \text{and} \quad v^i x_i >0 \ ,
\end{equation}
where $u_t = -W (\alpha - \beta_i v^i)$
is the time component of the fluid 4-velocity (with a lowered index),
$\alpha$ is the lapse, $\beta^i$ is the shift vector,
$W$ is the Lorentz factor, and $x^i = (x,y,z)$.
For Eq.~\eqref{eq:unbound} we assume that the fluid elements follow geodesics
and requires that the orbit is unbound and has an outward pointing velocity,
cf.~also~\cite{East:2012ww}.
As pointed out in~\cite{Dietrich:2016hky} a possible drawback is that material
which gets ejected and decompresses
can obtain densities of the order of the artificial atmosphere, which is
added to allow stable GRHD simulations.
Once the density drops below the atmosphere value material is
not included in the ejecta computation anymore and the ejecta mass is
possibly underestimated.
For the case where we compute the ejecta mass due to the matter flux across a
coordinate sphere this effect is reduced, since the decompression of
material outside the coordinate sphere does not influence the ejecta mass computation.
Specifically, based on the continuity equation and the Gauss theorem we compute the unbound mass as:
\begin{equation}
M_{\rm ej}^{\mathcal{S}} = 
{\int_0}^t dt^{\prime} \int_{r=r_{\mathcal{S}}} [ D_u \left(\alpha v^i +\beta^i\right) n_i ] r^2 d\Omega  ,
\label{eq:Mejecta_S}
\end{equation}
where $n_i = x_i/r$ with $r = \sqrt{x^ix_i}$. $D_u$ denotes the unbound fraction of
conserved rest mass density $D = W\rho$. The estimates $M_{\rm ej}^{\mathcal{V}}$ and 
$M_{\rm ej}^{\mathcal{S}}$ are compared in Sec.~\ref{subsec:ejecta}. \\

\paragraph*{\textbf{Spectrograms.}} 
As discussed in the introduction, density oscillations can be induced in the stars during
close encounters on highly eccentric orbits.
Most of the oscillation energy is released in the \textit{f}-modes
which imprint their own characteristic GWs on top of the
GWs generated by the orbital motion. In order to study
these \textit{f}-mode oscillations, we consider the spectrograms of the individual,
$\l=2$, dominant modes of the curvature scalar $r\Psi _4$.

We compute the frequency domain GW signal $r\tilde{\Psi}^{\lm}_4 (f)$
using the discrete Fourier transform implemented in MATLAB for time
intervals of length $t_2 - t_1$.
The corresponding power spectral density (PSD) is given by 
\begin{equation}
\PSD(r\Psi _4 ^{\lm})(f) = (t_2 - t_1)^2 \ \left\lvert r\tilde{\Psi} _4 ^{\lm}(f)\right\rvert ^2
\label{eq:PSD}
\end{equation}
to get the distribution of power into frequency components. This gives us a
time series distribution of power and frequency, disentangling the dynamics
of the system. From the spectrograms we find the frequency at which the NSs
oscillate and also other frequencies of the dominant dynamics like inspiral,
merger\footnote{Note that we define the moment of merger as the peak in the amplitude
of the GW strain, $rh_{22}$.} and the oscillations of the merger remnant. We also compute
separate PSDs for the premerger and postmerger signal. 
These results will be discussed in more detail 
in Sec.~\ref{sec:GWs}.\\

\paragraph*{\textbf{Removing displacement-induced mode mixing from the GW signal:-}}
Since we are considering symmetric systems, the $\textit{f}$-mode oscillations seen
in the waveform will be twice those of an individual star. 
However, because the
stars are not located at the origin, there will be some mode mixing which can not be neglected.
Thus, it is not possible to simply divide the $\textit{f}$-mode contribution to a given mode by 
$2$ to obtain the contribution from an individual star
that would be extracted if it was at rest at the origin.

To obtain approximate expressions for an individual star's multipole moments 
(as would be extracted with it at the origin), 
we will consider times when the stars are well-separated and 
approximately separate the $\textit{f}$-mode signal from the 
signal from the orbit by taking a moving average of 
$\Psi^4_{2, \pm 2}$ with a window width given by the period 
of the $f$-mode.\footnote{Here and in all other mode mixing analysis, we denote
the $(\ell, m)$ mode of $\Psi_4$ by $\Psi^4_{\ell, m}$, for notational simplicity.} Specifically,
we define 
$\Psi^{4, f\text{-mode}}_{2, \pm 2} := \Psi^4_{2, \pm 2} - \Psi^{4, \text{avg}}_{2, \pm 2}$, 
where $\Psi^{4, \text{avg}}_{2, \pm 2}$ denotes the result
of applying the moving average to $\Psi^4_{2, \pm 2}$.

Additionally, since the decay times of the $f$-mode 
oscillations are much longer than the time between periastra, 
one can treat the $\textit{f}$-mode signal from a single star in a given $(\ell_\text{i}, m_\text{i})$
mode as a simple sinusoid, $\Psi^{4, f\text{-mode}}_{\ell_\text{i}, m_\text{i}} e^{i\omega_{f\text{-mode}} t}$, and can compute
the mode mixing due to the stars' displacement from the origin 
analytically using Eq.~(43b) in~\cite{Boyle:2015nqa} (summing the series to obtain an exponential).
This mode mixing is due to the variations in the retarded time at different points on the extraction sphere; 
we compute the retarded time approximately using the coordinate tracks of the stars.
We thus take the retarded time to be $t_{\text{ret}, 0} + \alpha(\theta,\phi)$, 
where $t_{\text{ret}, 0}$ denotes the retarded time for a source at the origin, and $\alpha(\theta,\phi) = d_1\sin\theta\cos(\phi-\Upsilon)$ 
[cf.\ the discussion above Eq.~(4) in~\cite{Boyle:2015nqa}].
Here $\theta,\phi$ are the angular spherical coordinates on the 
extraction surface and $d_1(\cos\Upsilon, \sin\Upsilon,0)$
is the coordinate track of star~1; the negative of this gives the track
of the star's companion.\footnote{Here we are assuming that the binary's center-of-mass (COM) is at the origin, for simplicity of exposition. The COM actually drifts over the
course of evolution, and its displacement from the origin is considerable in a few cases (increasing with decreasing eccentricity). We will discuss later how to account for this drift in the mode mixing analysis.} Additionally, we take
each star to only have an intrinsic $(\ell_\text{i}, m_\text{i})$ mode, 
since we can include further modes by linearity.
Thus, the $\textit{f}$-mode contribution to the (spin-weighted) 
spherical harmonic modes extracted from the evolution of the binary will be given by
\begin{equation}
\begin{split}
\Psi^{4, \text{ext}}_{\ell, m} &= 2\Psi^{4, f\text{-mode}}_{\ell_\text{i}, m_\text{i}}\int_{S^2} \cos[d_1\omega_{f\text{-mode}}\sin\theta\cos(\phi-\Upsilon)]\\
&\quad \times {}_{-2}Y_{\ell_\text{i}, m_\text{i}}(\theta,\phi){}_{-2}Y^*_{\ell, m}(\theta,\phi) d\Omega\\
&=: \mu_{\ell_\text{i}, m_\text{i}; \ell, m}\Psi^{4, f\text{-mode}}_{\ell_\text{i}, m_\text{i}},
\end{split}
\end{equation}
where ${}_{-2}Y_{\ell m}$ is the spin-$(-2)$-weighted spherical harmonic, the star denotes the complex conjugate, and we have defined the mode mixing coefficient $\mu_{\ell_\text{i}, m_\text{i}; \ell, m}$, which describes
mixing from the $(\ell_\text{i}, m_\text{i})$ intrinsic mode of an individual star into the $(\ell, m)$ mode of the binary. Note that we have
$\mu_{\ell_\text{i}, m_\text{i}; \ell, m} = \mu^*_{\ell, m; \ell_\text{i}, m_\text{i}}$.

Defining $\zeta := d_1\omega_{f\text{-mode}}$, we obtain mode mixing coefficients of
\begin{subequations}
\begin{align}
\mu_{2, \pm 2;2,\pm 2} &= \frac{5}{8}\cS(-3,-3;\zeta)\nonumber\\
&= 2 - \frac{5}{21}\zeta^2+ O(\zeta^4),\\
\mu_{2,0;2,0} &= \frac{15}{4}\cS(-3,1;\zeta)\nonumber\\
&= 2 - \frac{3}{7}\zeta^2 + O(\zeta^4),\\
\mu_{2, \pm 2;2,0} &= \frac{5e^{\pm 2i\Upsilon}}{4}\sqrt{\frac{3}{2}}\cS(5,1;\zeta)\nonumber\\
&= -\frac{2e^{\pm 2i\Upsilon}}{7\sqrt{6}}\zeta^2 + O(\zeta^4),\\
\mu_{2, \pm 2;2,\mp 2} &= \frac{5e^{\pm 4i\Upsilon}}{8}\cS(-35,5;\zeta)\nonumber\\
&= \frac{e^{\pm 4i\Upsilon}}{1512}\zeta^4 + O(\zeta^6),
\end{align}
\end{subequations}
where
\begin{equation}
\begin{split}
\cS(a,b;z) &:= \frac{z(3a + 2b z^2)\cos z + [-3a + (a - 2b) z^2 + z^4]\sin z}{z^5}\\
&\;= -a\frac{j_2(z)}{z^2} - 2b\frac{j_1(z)}{z} + j_0(z)
\end{split}
\end{equation}
is a sinc-like function; we also give the expression in terms of the spherical Bessel functions of the first kind, $j_\ell$.
We give the first few terms of the power series expansions 
to give intuition about the behavior of the mixing coefficients for small $\zeta$.
We do not give the mixing coefficients involving the $(2,\pm 1)$ 
modes, as they vanish, due to the symmetry of the system (since we are assuming that the binary's center-of-mass is at the origin in
the current discussion). While the $\mu_{2, \pm 2;2,\mp 2}$
coefficients do not vanish, they are suppressed by higher 
powers of $\zeta$, only starting at $O(\zeta^4)$, and are 
small enough for the situations
we are considering (magnitudes $< 0.03$) such that we will simply ignore them and 
focus on the much larger effects from the other mixing coefficients given above. Similarly,
we do not consider mixing of higher-$\ell$ modes into the 
$\l = 2$ modes, since we expect those modes to have intrinsically smaller amplitudes, compared with the $\l = 2$ modes.

We also do not consider mode mixing due to the boosts, 
since the speeds due to the binary's orbital motion are relatively small 
($< 0.15$) in the region between bursts, where the stars are
well separated and the contributions from the mode mixing due 
to displacement from the origin are largest: The linear-in-velocity contribution to the mode
mixing due to the boost vanishes, due to the symmetry of the binary 
(see~\cite{Gualtieri:2008ux} for explicit expressions in the linearized case; the general computation
is discussed in~\cite{Boyle:2015nqa}).

If one needs to include the displacement of the binary's center-of-mass from the origin, then it is simple to convert the expressions above
to cover the general case: The contribution from a single star is just half of the total, so if the two stars are located at $d_1(\cos\Upsilon_1, \sin\Upsilon_1,0)$
and $-d_2(\cos\Upsilon_2, \sin\Upsilon_2,0)$, respectively, then, defining $\zeta_A := d_A\omega_{f\text{-mode}}$ with $A\in\{1,2\}$, we have, e.g.,
\begin{equation}
\begin{split}
\mu_{2, \pm 2;2,0}^\text{general} &= \frac{5}{8}\sqrt{\frac{3}{2}}\sum_{A=1}^2 e^{\pm 2i\Upsilon_A}\cS(5,1;\zeta_A).
\end{split}
\end{equation}
We use these general expressions for all the results presented in the paper, though the simpler expressions with the center-of-mass at the origin suffice in almost
all cases. Additionally, for completeness, we also give the mixing from the intrinsic
$(2,\pm 2)$ modes into the $(2,\pm 1)$ modes, though we do not consider these further in this paper:
\begin{subequations}
\begin{align}
\mu_{2,\pm 2;2,\pm 1}^\text{general} &= \frac{5i}{8}\sum_{A=1}^2 (-1)^A e^{\pm i\Upsilon_A}\cT(3,1;\zeta_A)\nonumber\\
&= \frac{i}{3}(\zeta_1e^{\pm i\Upsilon_1} - \zeta_2e^{\pm i\Upsilon_2}) + O(\zeta_{1,2}^3),\\
\mu_{2,\pm 2;2,\mp 1}^\text{general} &= \frac{5i}{8}\sum_{A=1}^2 (-1)^A e^{\pm 3i\Upsilon_A}\cT(-5,-3;\zeta_A)\nonumber\\
&= -\frac{i}{168}(\zeta_1^3e^{\pm 3i\Upsilon_1} - \zeta_2^3e^{\pm 3i\Upsilon_2}) + O(\zeta_{1,2}^5),
\end{align}
\end{subequations}
where
\begin{equation}
\begin{split}
\cT(a,b;z) &:= \frac{z(3a + z^2)\cos z + (-3a + 2bz^2)\sin z}{z^4}\\
&\;= -3a\frac{j_1(z)}{z^2} + 2b\frac{j_0(z)}{z} - y_0(z).
\end{split}
\end{equation}
Here $y_\ell$ are the spherical Bessel functions of the second kind.
[Note that Mathematica (at least as of version~11) will not evaluate the $\theta$ integral giving the second of these mixing coefficients as is. However, if one uses the Maclaurin series for the Bessel function one obtains upon performing the $\phi$ integral first, and then integrates term-by-term, Mathematica will sum the resulting infinite series with no problems.]

\section{Configurations}
\label{sec:config}

In total, we consider $14$ different physical configurations, summarized in
Table~\ref{tab:config}. All setups employ at least the R2 resolutions,
cf.~Table~\ref{tab:grid}.
A subset of configurations is also simulated with grid setups R1, R3, and R4.
We mark those setups with an asterisks ($^*$) in Table~\ref{tab:config}.
The name of the simulations refer to:
the EOS, the input eccentricity parameter used in Eq.~\eqref{eq:KV},
and the spin orientations. All results shown in the paper are obtained from the
R2 resolution simulations, unless otherwise noted.

For this work we decided to focus on equal-mass setups with baryonic
masses $M^A _b=M^B _b\simeq 1.504M_\odot$. The stars are either
irrotational or have dimensionless spins of $\chi \simeq 0.1$ oriented parallel to
the orbital angular momentum. To compute the rotation frequency of the stars
corresponding to this dimensionless spin, we compared our results
for the SLy EOS against rigidly rotating neutron stars computed with the publicly
available \texttt{Nrotstar} module of the LORENE library~\cite{LoreneCode}.
Such a comparison is valid since rotating stars constructed employing the constant
rotational velocity approach, as in SGRID~\cite{Tichy:2011gw},
have an almost zero shear~\cite{Dietrich:2015pxa,Tichy:2012rp}.
We obtain for the SLy setups a rotational frequency of $f\sim 191{\rm Hz}$. 

In addition to the definition of eccentricity given in Sec.~\ref{ssec:initial} 
and the references there, we also compute the post-Newtonian (PN) eccentricity
from the Arnowitt-Deser-Misner (ADM) expressions for the energy and angular momentum.
In~\cite{Dietrich:2015pxa} an extensive comparison was performed between different
order PN eccentricities and the eccentricity measure used in the helliptical symmetry
vector in SGRID.
Following this work, we use the 3PN expression for eccentricity, Eq.~(4.8)
in~\cite{Dietrich:2015pxa}, computed following Mora and Will~\cite{Mora:2003wt}.
For completeness we also give the full expression below:

\begin{equation}
\begin{split}
\hat{e}_{\rm 3PN} ^2 &= 1-2 \xi + \left[-4-2 \nu + (-1 + 3 \nu) \xi \right] E_b \\
&\quad + \left[\frac{20-23 \nu}{\xi} -22 + 60 \nu + 3 \nu ^2 - (31 \nu + 4 \nu ^2) \xi \right] E_b ^2 \\
&\quad + \left[\frac{-2016+(5644-123 \pi ^2) \nu - 252 \nu ^2}{12 \xi ^2} \right. \\
&\quad + \left. \frac{4848 + (-21128 + 369 \pi ^2) \nu + 2988 \nu ^2}{24 \xi} \right. \\
&\quad - \left. 20 + 298 \nu - 186 \nu ^2 - 4 \nu ^3 \right. \\
&\quad + \left. \left(-30 \nu + \frac{283}{4} \nu ^2 + 5 \nu ^3\right) \xi \right]   E_b ^3 .
\label{eq:3PNecc}
\end{split}
\end{equation}
Here $\xi := - E_b \ell ^2$ and $\nu := M^AM^B/M^2$ is the symmetric mass ratio,
where $E_b = (M_\text{ADM}/M - 1)/\nu$ is the binary's (reduced) binding energy,
$\ell = (J_\text{ADM} - S^A - S^B)/(M^2 \nu)$ is its specific orbital angular momentum,
$M^{A,B}$ are the individual gravitational masses of the stars in isolation,
$M := M^A + M^B$ is the binary's total mass, $M_\text{ADM}$ is its ADM mass,
$J_\text{ADM}$ is its ADM angular momentum, and $S^{A,B}$ are the
(dimensionful) spins of the stars.
The eccentricities input into SGRID and the computed 3PN
eccentricities are listed in Table~\ref{tab:config}
for all the configurations.

Note that Eq.~\eqref{eq:3PNecc} only includes nonspinning point mass contributions to the energy and angular momentum.
However, we have checked that including the leading spin-orbit and tidal contributions gives almost identical results, only
changing the final digit of the value we quote by $1$ in one case. The ingredients to perform these computations
are given in Eqs.~(3.2), (3.6), (3.15), and (3.16) in Mora and Will~\cite{Mora:2003wt}, and yield contributions to
$\hat{e}_{\rm 3PN} ^2$ of
\begin{equation}
\begin{split}
&+4\left[\left(\frac{5}{\xi^{1/2}} - 3\xi^{1/2}\right)\tilde{S} + \frac{\delta M}{M}\left(\frac{2}{\xi^{1/2}} - \xi^{1/2}\right)\tilde{\Delta}\right](-E_b)^{3/2}\\
&- 8\left(\frac{4}{\xi^4} - \frac{10}{\xi^3} + \frac{5}{\xi^2}\right)\kappa_2^T E_b^5,
\end{split}
\end{equation}
where $\tilde{S} := (S^A + S^B)/M^2$, $\tilde{\Delta} := (S^B/M^B - S^A/M^A)/M$, $\delta M := M^A - M^B$,
and $\kappa_2^T := 2(R^A/M)^5(M^B/M^A)k_2^A + (A \leftrightarrow B)$ is the tidal coupling constant introduced in~\cite{Damour:2009wj}, where $R^A$ and $k_2^A$
are the areal radius and quadrupolar dimensionless tidal Love number of star~$A$, respectively. Including the spin term increases the eccentricity estimate for MS1b$-e0.50 ^{\uparrow\uparrow}$ from
$0.53$ to $0.54$, due to rounding; the tidal term has a negligible effect. This is the only case for which adding either of these terms affects the eccentricity we quote. However,
these results suggest that a higher-order calculation of the spin-dependent contributions, going beyond the ingredients provided by Mora and Will, will be necessary to make an
accurate PN eccentricity estimate for more highly spinning cases.

\section{Dynamics}
\label{sec:dyn}

\subsection{Qualitative discussion}

\begin{figure}[t]
\includegraphics[width=0.5\textwidth]{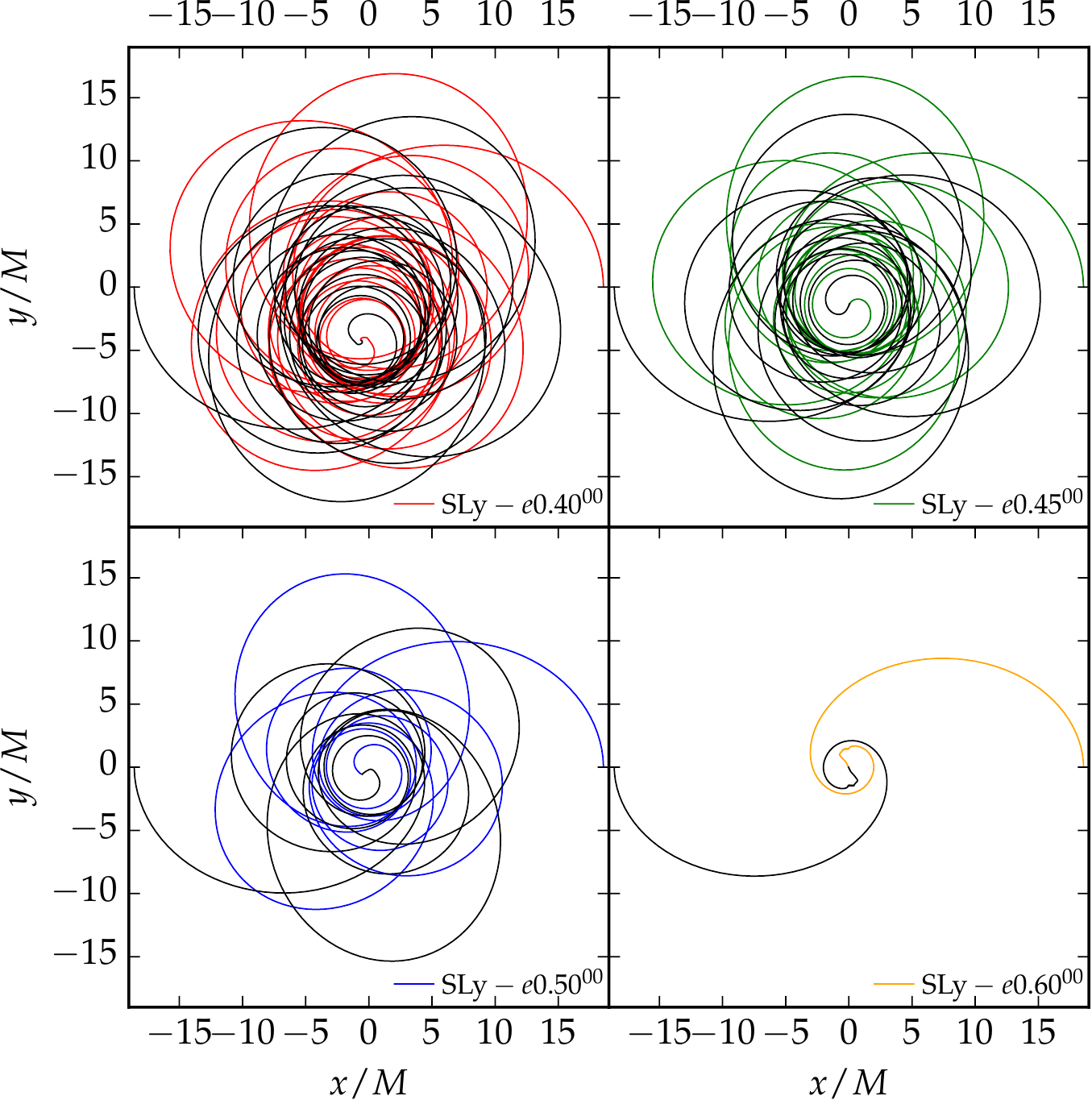}
\caption{
Tracks of the star centers for the SLy$^{00}$ cases.
Each panel refers to a different input eccentricity:
$e=0.40$ (upper left, red), $e=0.45$ (upper right, green),
$e=0.50$ (lower left, blue), $e=0.60$ (lower right, orange).
The star that originally starts on the left-hand side always has its
track colored black.}
\label{fig:tracks:SLy00}
\end{figure}

\begin{figure}[t]
\includegraphics[width=0.5\textwidth]{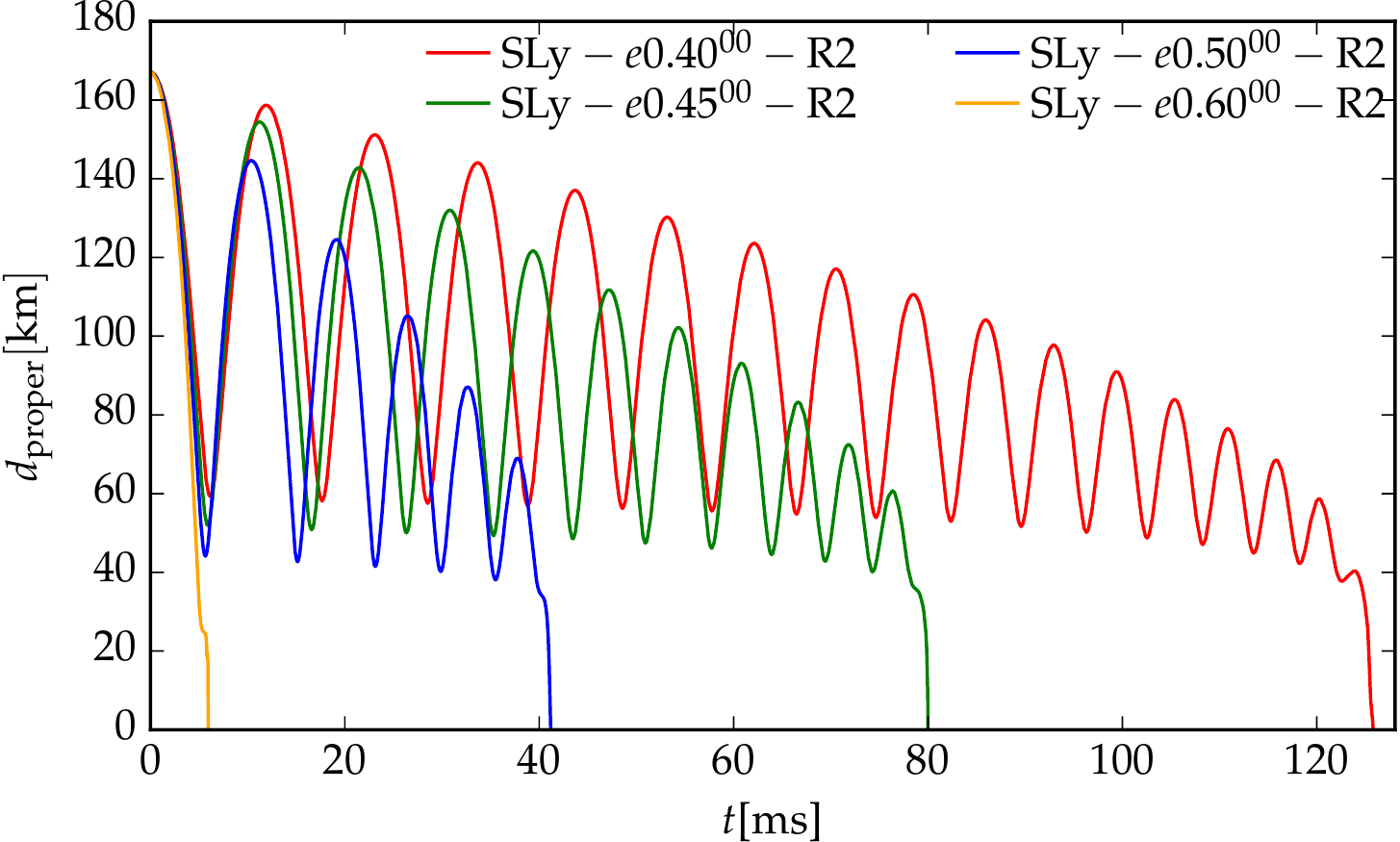}
\caption{
The evolution of the proper distance between the two non-spinning NSs 
with SLy EOS and with different initial eccentricities. For the system with 
less initial eccentricity, the stars do not approach each other as closely, compared to the 
system with relatively larger initial eccentricity. In other words, the 
larger the initial eccentricity the smaller is the impact parameter.}
\label{fig:SLy:properdistance}
\end{figure}

\begin{figure*}[t]
\centering
\includegraphics[width=0.8\textwidth]{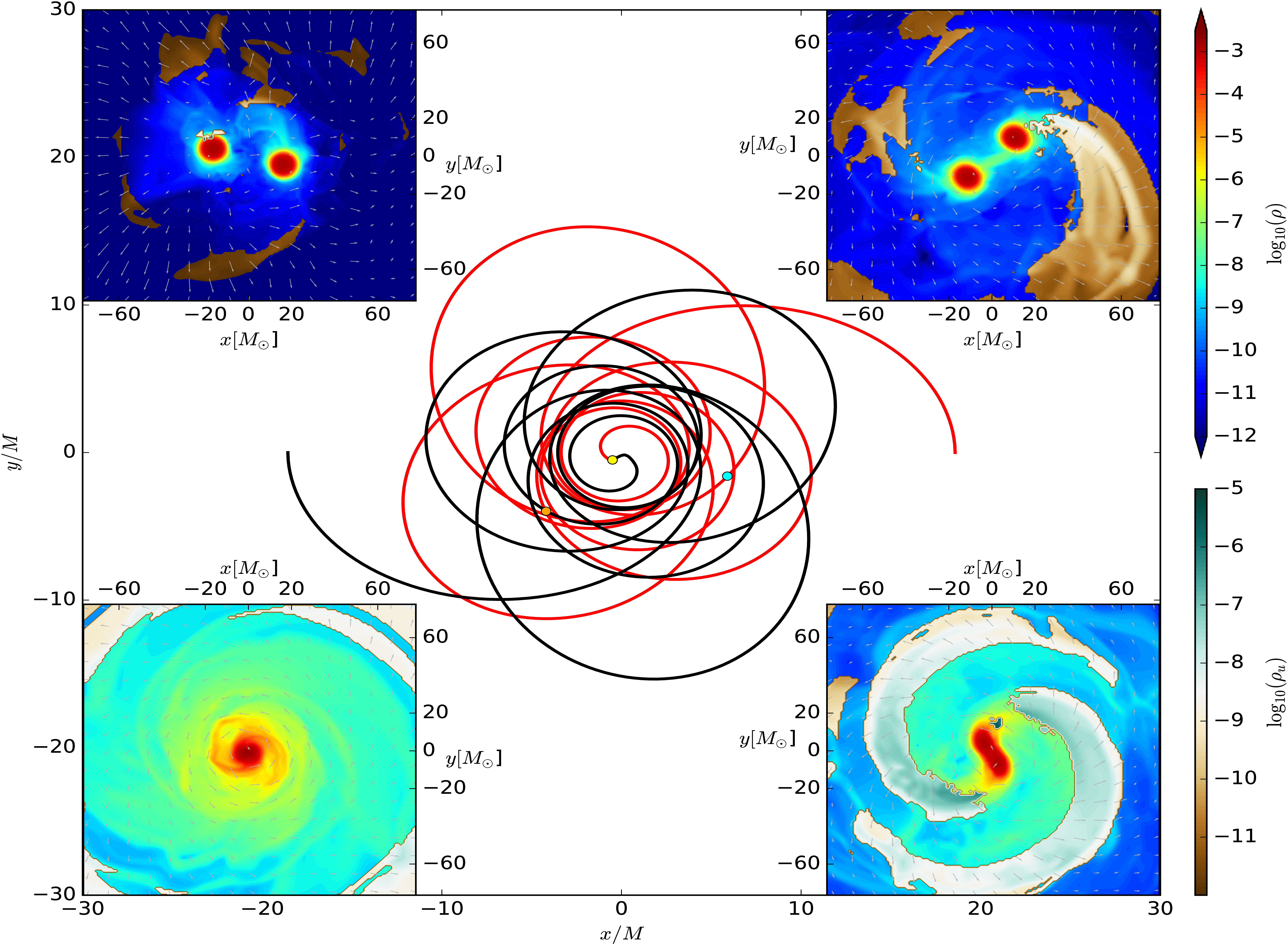}
\caption{
NS trajectories computed from the coordinate positions of the lapse
minima for the two stars
of the SLy-e$0.50^{00}-$R2 configuration.
We also show 2d-plots of the density and velocity field
at different times, with the unbound material [computed using the criterion in Eq.~\eqref{eq:unbound}] shown in the brown
to green color scale, while the bound material is shown in a blue to red color scale.
The times corresponding to the inset plots are marked with colored circles on the trajectory.
\textit{Top left (cyan):}
Time after the first encounter, where the stars come as close as
$\sim$45~km (proper distance).
As the stars undergo a tightly bound whirl orbit, one can see that some low-density matter is ejected.
\textit{Top right (orange):}
After periastron encounter at a later time in the evolution.
Here also we see similar features as in the \textit{top left} plot except that
there is also mass transfer at a density $\sim
\mathcal{O}(10^{-9})-\mathcal{O}(10^{-8})$
[$\mathcal{O}(10^{8})-\mathcal{O}(10^{9})$~g/cm$^3$].
\textit{Bottom right (yellow):} 
Time just before the merger. 
Unbound matter with densities
$\mathcal{O}(10^{-7})-\mathcal{O}(10^{-6})$
[$\mathcal{O}(10^{10})-\mathcal{O}(10^{11})$~g/cm$^3$]
is ejected from the tidal tail and as the unbound matter expands its
density decreases.
\textit{Bottom left (yellow):} 
Post-merger phase when the cores of the NSs have merged and formed a
hypermassive NS.}
\label{fig:2dtracks:SLy00}
\end{figure*}
The simulations are performed with configurations
employing $3$ to $4$ different input eccentricities.
One finds that although the initial distance is the same for all
configurations,
the number of orbits until merger varies significantly as visible in
Fig.~\ref{fig:tracks:SLy00}. In particular, for
an increasing eccentricity, one finds the number
of orbits to be $\sim15.5, 10.5, 5.5$, and $0.5$ as
computed from Fig.~\ref{fig:SLy:properdistance}.
The orbits are seen to undergo apsidal (orbital) 
precession, where the orbit of the NSs rotates in the plane of motion.
The reduction of eccentricity due to the emission of
GWs~\cite{Peters:1964zz} is clearly visible from the evolution
of the proper distance between the neutron
stars as in Fig.~\ref{fig:SLy:properdistance}.
As an example, we present the
SLy$-e0.50^{00}-$R2 configuration in Fig.~\ref{fig:2dtracks:SLy00}.
For this case the stars perform $\sim 5.5$ encounters until
they finally merge.
During the encounters of the NSs, when they come within $\sim 45$~km
of each other, the deformation of the
individual stars increases due to the stronger tidal forces
induced by the companion. Because of these deformations the stars start
to oscillate.
Furthermore, at later a stage in the evolution, a fraction
of the material is ejected from the system during grazing encounters.
We show the ejecta on a brown
to green color scale in Fig.~\ref{fig:2dtracks:SLy00} where the bound density
is shown from blue to red. In particular, the upper right inset shows a large
amount of matter which gets ejected from the system just after a grazing
periastron encounter. At the merger (lower right panel) one clearly sees tidal tails
behind the two stars from which most material is released.
Finally, after the merger, the remnant stabilizes and either forms a stable
massive neutron star (MNS, MS1b cases) or a hypermassive neutron star (HMNS, SLy cases).

\subsection{Energetics}

\begin{figure}[t]
\includegraphics[width=0.5\textwidth]{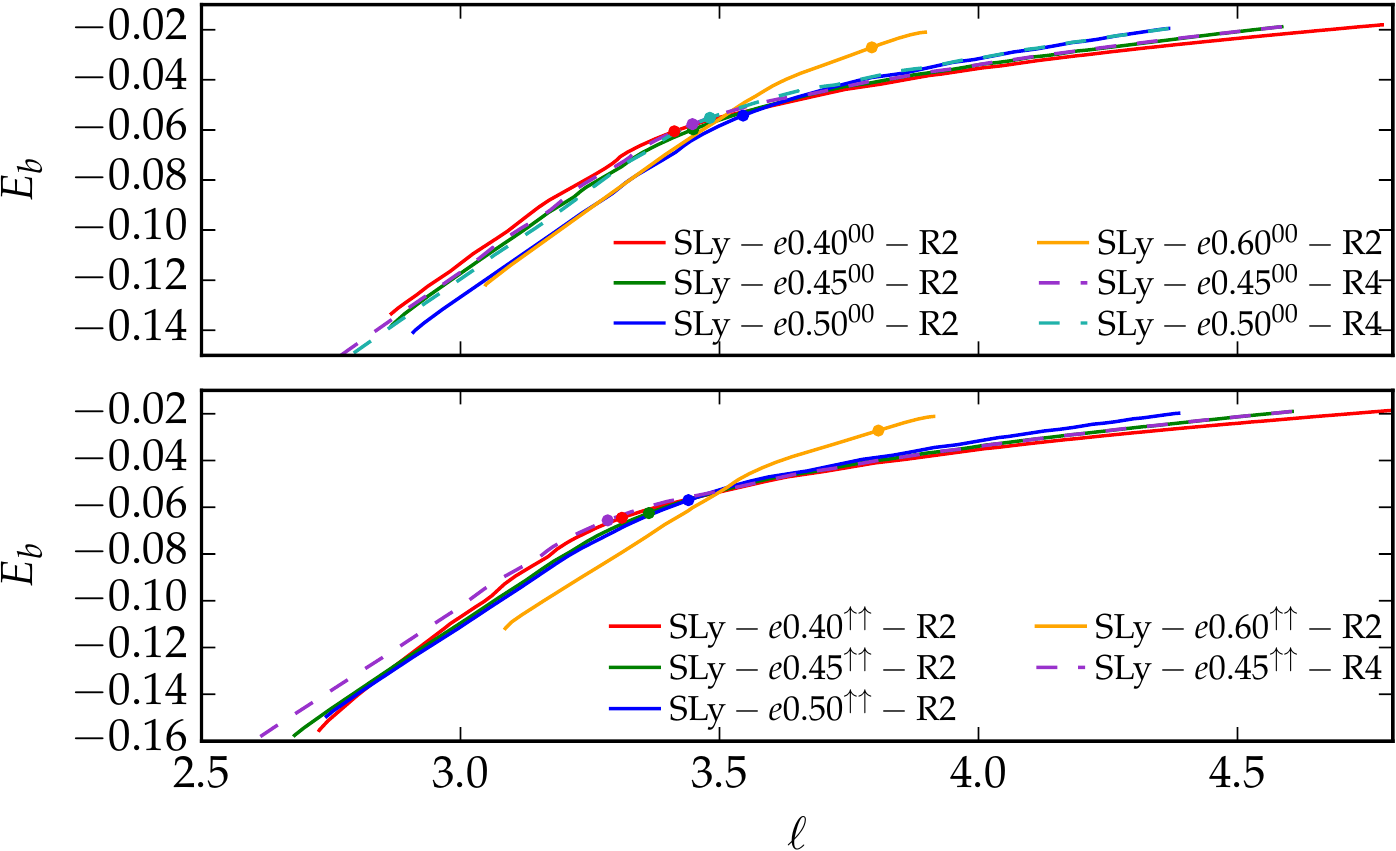}
\caption{
Binding energy vs.~specific angular momentum curves $E_b(\ell)$ for SLy EOS
for nonspinning (top panel) and aligned spin configurations (bottom panel).
The moments of merger are marked by circles.
}
\label{fig:Ej:SLy}
\end{figure}

\begin{figure}[t]
\includegraphics[width=0.5\textwidth]{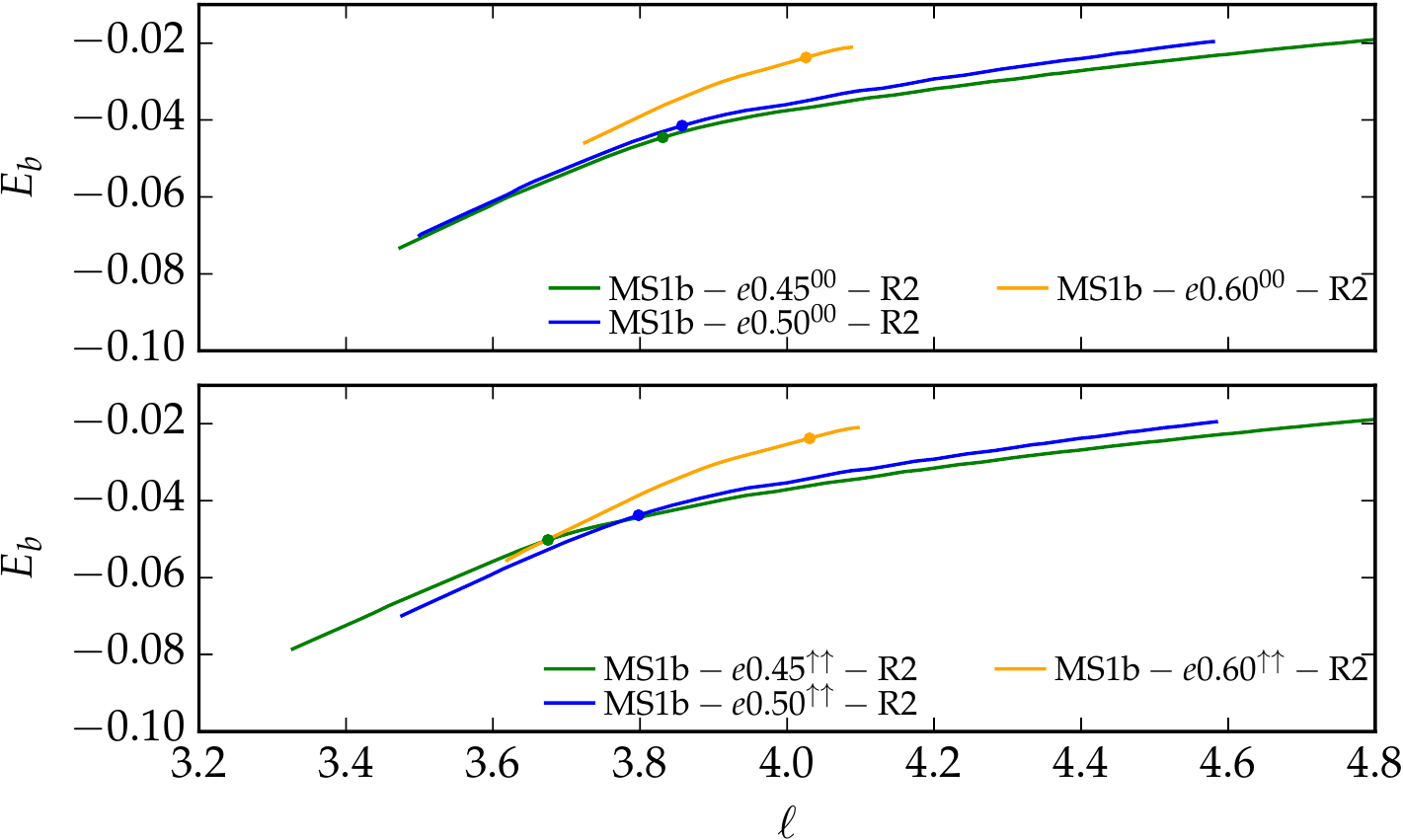}
\caption{
Binding energy vs.~specific angular momentum curves $E_b(\ell)$ for MS1b EOS
for nonspinning (top panel) and aligned spin configurations (bottom panel). 
The moments of merger are marked by circles.
}
\label{fig:Ej:MS1b}
\end{figure}

For a qualitative discussion of the conservative dynamics of the system, 
we compute the binding energy $E_b$ and specific angular momentum 
$\ell$ from our numerical simulations, see, e.g., \cite{Damour:2011fu,Bernuzzi:2013rza,Dietrich:2017feu}.

As examples, we show in Fig.~\ref{fig:Ej:SLy}
and Fig.~\ref{fig:Ej:MS1b} the binding
energy for the nonspinning (top panel) and spin-aligned (bottom panel)
SLy and MS1b configurations respectively.
For increasing eccentricity, we find that the curves start
with less angular momentum, even though the initial distance is fixed.
In the limit of $e\nearrow1$, one would obtain
an initial specific orbital angular momentum of $\ell=0$, because then,
at the star center, the symmetry vector of Eq.~(\ref{eq:KV})
and thus the fluid velocity have no component perpendicular to the 
position vector of the star's center (i.e., no $y$-component).

During the inspiral, the system emits energy and angular momentum 
in the form of GWs.
In general, for an increasing eccentricity, the slope of the
individual curves increases.
Stated differently, the dimensionless frequency
$M\hat{\Omega} = \partial E_b/\partial\ell $
at a given specific orbital angular momentum
is larger for increasing values of the eccentricity.
The systems are less bound for larger eccentricities at
fixed $\ell$ during the early part of the inspiral as predicted by PN 
theory; see Ref.~\cite{Mora:2003wt}. The behavior of the dimensionless
frequency with eccentricity is opposite to that predicted by PN theory. The
cause of the difference is not clear, but may be due to the fact that most of the
energy and angular momentum are lost near periastron, where the PN approximation
is not very accurate.
In the most extreme case the stars
perform a head on collision for which the angular momentum
remains zero regardless of the separation.
The stars can be almost unbound even for very small (or even zero) angular momentum. This
cannot be achieved for setups with decreasing eccentricity.

It is important to notice that close to the merger and after the merger 
the ordering of the binding energy curves changes. The merger itself is marked 
by circles in Fig.~\ref{fig:Ej:SLy}. This observation can be understood
as follows.
The moment of merger marks the initial configuration for the evolution 
of the postmerger. Consequently when the angular momentum and binding
energy is larger
the remnant is less bound and rotates faster, i.e., $\hat{\Omega}$ is larger. 

Similar results are obtained for the MS1b EOS: See
Fig.~\ref{fig:Ej:MS1b}.

\subsection{Merger remnant}
\label{sec:PM}

\begin{table}[tp]
\caption{Properties of the merger remnant for the SLy cases.
We do not report simulations for MS1b since all these 
setups form a stable MNS. The columns represent:
(i) the name of the configuration
(ii) the lifetime, $\tau$, of the HMNS formed during our simulation,
given in $M_{\odot}$ and in {\rm ms};
(iii) the final mass of the BH, $M_\text{BH}$, if the HMNS collapsed during
our simulation;
the dimensionless spin of the final BH, $\chi_\text{BH}$
and the mass of the disk surrounding the BH, $M_\text{disk}$.
We also show the quantities computed from different resolutions for
some configurations where they are available.
The quantities for configurations that did not undergo
collapse to a BH are marked with ``$-$''.}
\label{tab:remnant} 
\centering
\begin{small}
\begin{tabular}{c|cc|ccc}
Name & \multicolumn{2}{c|}{$\tau$} & $M_\text{BH}$ & $\chi_\text{BH}$ & $M_\text{disk}$ \\
     & \begin{footnotesize} $[M_{\odot}]$  \end{footnotesize}& \begin{footnotesize} $[\rm ms]$  \end{footnotesize}& \begin{footnotesize} $[M_{\odot}]$  \end{footnotesize}&                  & \begin{footnotesize} $[M_{\odot}]$ \end{footnotesize}\\
\toprule
SLy$-e0.40^{00}-{\rm R2}$ &  $1300$& $6.4$ & $2.58$ & $0.62$ & $0.08$\\
\hline
SLy$-e0.45^{00}-{\rm R1}$ &  $2100$& $10.3$ & $2.53$ & $0.55$ & $0.09$\\

SLy$-e0.45^{00}-{\rm R2}$ &  $6000$& $29.6$ & $2.45$ & $0.48$ & $0.13$\\

SLy$-e0.45^{00}-{\rm R3}$ &  $>11150$ & $>54.9$ & $-$ & $-$ & $-$\\

SLy$-e0.45^{00}-{\rm R4}$ &  $3200$& $15.8$ & $2.48$ & $0.54$ & $0.09$\\
\hline
SLy$-e0.50^{00}-{\rm R1}$ &  $>6820$ & $>33.6$ & $-$ & $-$ & $-$\\

SLy$-e0.50^{00}-{\rm R2}$ &  $>9085$ & $>44.7$ & $-$ & $-$ & $-$\\

SLy$-e0.50^{00}-{\rm R3}$ &  $1700$& $8.4$ & $2.55$ & $0.59$ & $0.07$\\

SLy$-e0.50^{00}-{\rm R4}$ &  $2700$& $13.3$ & $2.51$ & $0.56$ & $0.09$\\
\hline
SLy$-e0.60^{00}-{\rm R2}$ &  $>13100$ & $>64.5$ & $-$ & $-$ & $-$\\
\hline
SLy$-e0.40^{\uparrow\uparrow}-{\rm R2}$ &  $1900$& $9.4$ & $2.56$ & $0.60$ & $0.07$\\
\hline
SLy$-e0.45^{\uparrow\uparrow}-{\rm R1}$ &  $5400$& $26.6$ & $2.42$ & $0.46$ & $0.16$\\

SLy$-e0.45^{\uparrow\uparrow}-{\rm R2}$ &  $2400$& $11.8$ & $2.54$ & $0.59$ & $0.07$\\

SLy$-e0.45^{\uparrow\uparrow}-{\rm R3}$ &  $2500$& $12.3$ & $2.54$ & $0.58$ & $0.07$\\

SLy$-e0.45^{\uparrow\uparrow}-{\rm R4}$ &  $>10000$ & $>49.3$ & $-$ & $-$ & $-$ \\
\hline
SLy$-e0.50^{\uparrow\uparrow}-{\rm R2}$ &  $>9440$ & $>46.5$ & $-$ & $-$ & $-$ \\
\hline
SLy$-e0.60^{\uparrow\uparrow}-{\rm R2}$ &  $>8931$ & $>44.0$ & $-$ & $-$ & $-$ \\
\hline
\end{tabular}
\end{small}
\end{table}

Due to the choice of the particular total mass of $M\sim 2.7 M_\odot$,
we find in general three different outcomes for our simulations.
One possible outcome is the formation of a stable MNS
in cases where the total mass of the remnant is below the maximum allowed
mass of a spherically symmetric star for the given EOS.
Since MS1b supports non-rotating stars with masses up to $\sim 2.76M_\odot$,
all remnants formed by the merger of the MS1b configurations are indeed MNSs.
The SLy EOS instead only supports nonrotating stars with masses up to $\sim 2.06M_\odot$.
Consequently the remnant is unstable and will collapse to a BH.
In fact, the total masses of the remnants formed during the merger of the SLy
setups even exceeds the maximum mass of $\sim 2.5M_\odot$  of a rigidly rotating SLy star,
which is the reason we characterize the remnants as HMNSs
~\cite{Baumgarte:1999cq,Hotokezaka:2013iia,Baiotti:2016qnr}.
Four out of the eight SLy configurations form a BH during the simulation time, at the R2 resolution; one additional configuration
only forms a BH during the simulation time at higher resolutions. (The other cases that do not form a BH during the simulation
time were not evolved at higher resolutions.)
We summarize the properties of the merger remnant for the SLy setups in Table~\ref{tab:remnant}.
In the following we discuss the lifetime
of the merger remnant, the properties of the final BH and the disk masses.

\paragraph*{\textbf{Lifetime of the merger remnant:-}}

The lifetime and merger properties in our simulations 
are mostly affected by the total mass of the system
and the EOS~\cite{Hotokezaka:2013iia,Dietrich:2015iva}. 
It is generally true that systems with a softer EOS collapse earlier
than systems with a stiffer EOS. In cases where the EOS is stiff enough, MNSs
can form, as in our MS1b simulations.
Systems with stiffer EOS are in general less bound at the merger
than systems with softer EOS (see Figs.~\ref{fig:Ej:SLy} and~\ref{fig:Ej:MS1b}).
The relations presented in~\cite{Read:2013zra,Bernuzzi:2014kca,Takami:2014tva} also imply that stiffer EOSs lead to
merger remnants with larger angular momentum support.
In addition to the angular momentum support, the pressure support in the
central regions is larger for stiffer EOS, and the merger remnant is even
further stabilized. Apart from the dependence on physical quantities,
e.g., angular momentum and EOS, the lifetime of the merger remnant is also
very sensitive to numerical errors and grid resolutions, as discussed in,
e.g., ~\cite{Hotokezaka:2013iia,Bernuzzi:2013rza,Bernuzzi:2015opx}. 
It is thus difficult to quantify the
collapse time, and uncertainties can be of the order of several milliseconds 
(see Table~\ref{tab:remnant}). When shocks form, e.g.~during the collision 
of the stars, even high order - high resolution shock capturing methods lose 
their high convergence properties~\cite{Bugner:2015gqa,Bernuzzi:2016pie,
Guercilena:2016fdl}.
We further find that the measurement of the remnant's lifetime is less robust 
for eccentric orbits than for quasicircular ones.
We think that this is due to the sensitive dependence of the postmerger evolution on
the number of close encounters before merger, which itself depends sensitively on
the eccentricity, spin, and/or resolution; see Appendix~\ref{app:conv}.

Despite these issues, a robust feature seems to be that
configurations with larger initial eccentricity with a fixed initial separation
have larger angular momentum at the moment of merger, as shown in Figs.~\ref{fig:Ej:SLy}
and~\ref{fig:Ej:MS1b}. 
Knowing that the angular momentum of a 
head-on collision is zero, this implies that 
there has to be some eccentricity value for which the angular 
momentum at merger reaches a maximum. 

Due to the larger angular momentum at the moment of merger, we 
expect a delayed BH formation. While we find that this is in agreement 
with the results at the R2-resolutions for which the largest set of 
simulations is available, higher resolutions suggest a more 
complicated picture. A further study to quantify the remnant lifetime 
is scheduled for the future. 
On the other hand, the imprint of spin is less clear.
While we find that spin aligned with the orbital
angular momentum leads to a delayed merger
(orbital hang-up effect~\cite{Campanelli:2006uy,Bernuzzi:2013rza}),
it is also seen that more angular momentum and energy
in the form of GWs is emitted before the merger
and hence the formed merger remnant has less angular
momentum leading to a faster collapse.

\paragraph*{\textbf{Black hole and disk properties:-}}
Four out of the 14 configurations collapse to a BH after the merger during
our simulation time at the R2 resolution, and one more collapses when run at
higher resolutions. We expect that if we evolved our configurations
with the SLy EOS for longer times then all configurations 
would have formed BHs.
In cases where a BH forms, we report the BH mass $M_\text{BH}$,
the dimensionless spin $\chi_\text{BH}$ of the BH,
and the mass of the accretion disk $M_\text{disk}$ in Table~\ref{tab:remnant}.

We find, independent of the exact setup, that systems with a larger 
lifetime form less massive black holes with generally smaller dimensionless spins, 
but more massive disks. However, no clear imprint of the eccentricity 
can be seen, taking into account the uncertainty of the numerical relativity simulations.

\section{Ejecta and EM Counterparts}
\label{sec:ejectaandEM}

\subsection{Ejecta}
\label{subsec:ejecta}

\begin{table}[tp]
\caption{Ejecta properties. The columns refer to: the name of the configuration, 
the mass of the ejecta from the volume integral $M_\text{ej}^{\mathcal{V}}$, 
the mass of the ejecta from the matter flux through coordinate spheres $M_\text{ej}^{\mathcal{S}}$, 
the kinetic energy of the ejecta $T_\text{ej}$, the $D$-weighted integral of $v^2=v^i v_i$ 
of fluid elements inside the orbital plane $\left\langle \bar{v} \right\rangle _{\rho}$ and 
perpendicular to it $\left\langle \bar{v} \right\rangle _{z}$.
We also show the quantities computed from different resolutions for 
some configurations.}
\label{tab:ejecta} 
\centering
\begin{small}
\begin{tabular}{c|cc|cc|cc}
Name & $M_\text{ej}^{\mathcal{V}}$& $M_\text{ej}^{\mathcal{S}}$ & \multicolumn{2}{c|}{$T_\text{ej}$}& $\left\langle \bar{v} \right\rangle _{\rho}$ & $\left\langle \bar{v} \right\rangle _{z}$ \\
       & \multicolumn{2}{c|}{\begin{footnotesize}$[10^{-2}M_{\odot}]$ \end{footnotesize} } & \begin{footnotesize} $[10^{-5}M_{\odot}]$ \end{footnotesize}& 
       \begin{footnotesize} [$10^{49}$ erg] \end{footnotesize} & \begin{footnotesize}$[c]$ \end{footnotesize}& \begin{footnotesize}$[c]$ \end{footnotesize}\\
\toprule
SLy$-e0.40^{00}-$R2 & $2.06$ & $1.96$ & $55.0$ & $98.3$ & $0.08$ & $0.15$ \\
\hline
SLy$-e0.45^{00}-$R1 & $2.00$ & $1.84$ & $14.0$ & $25.0$ & $0.09$ & $0.14$ \\
SLy$-e0.45^{00}-$R2 & $1.07$ & $1.02$ & $5.6$ & $10.0$ & $0.11$ & $0.14$ \\
SLy$-e0.45^{00}-$R3 & $2.03$ & $1.95$ & $10.0$ & $17.9$ & $0.13$ & $0.15$ \\
SLy$-e0.45^{00}-$R4 & $1.13$ & $1.08$ & $9.7$ & $17.4$ & $0.14$ & $0.16$ \\
\hline
SLy$-e0.50^{00}-$R1 & $0.90$ & $0.89$ & $1.0$ & $1.7$ & $0.06$ & $0.06$ \\
SLy$-e0.50^{00}-$R2 & $0.49$ & $0.46$ & $1.6$ & $2.9$ & $0.10$ & $0.11$ \\
SLy$-e0.50^{00}-$R3 & $1.90$ & $1.74$ & $20.0$ & $35.7$ & $0.09$ & $0.09$\\
SLy$-e0.50^{00}-$R4 & $0.32$ & $0.27$ & $1.5$ & $2.7$ & $0.11$ & $0.13$\\
\hline
SLy$-e0.60^{00}-$R2 & $0.23$ & $0.21$ & $1.3$ & $2.4$ & $0.12$ & $0.14$ \\
\toprule
SLy$-e0.40^{\uparrow\uparrow}-$R2 & $1.54$ & $1.52$ & $9.6$ & $17.2$ & $0.11$ & $0.15$ \\
\hline
SLy$-e0.45^{\uparrow\uparrow}-$R1 & $0.94$ & $0.90$ & $3.1$ & $5.5$ & $0.09$ & $0.11$ \\
SLy$-e0.45^{\uparrow\uparrow}-$R2 & $0.30$ & $0.35$ & $1.2$ & $2.2$ & $0.10$ & $0.11$ \\
SLy$-e0.45^{\uparrow\uparrow}-$R3 & $0.82$ & $0.81$ & $6.6$ & $11.7$ & $0.14$ & $0.15$\\
SLy$-e0.45^{\uparrow\uparrow}-$R4 & $0.68$ & $0.65$ & $2.9$ & $5.18$ & $0.10$ & $0.11$ \\
\hline
SLy$-e0.50^{\uparrow\uparrow}-$R2 & $0.70$ & $0.65$ & $1.7$ & $3.1$ & $0.08$ & $0.09$ \\
\hline
SLy$-e0.60^{\uparrow\uparrow}-$R2 & $1.15$ & $1.06$ & $4.4$ & $7.9$ & $0.11$ & $0.12$ \\
\toprule
MS1b$-e0.45^{00}-$R2 & $1.12$ & $1.05$ & $2.3$ & $4.1$ & $0.06$ & $0.08$ \\
\hline
MS1b$-e0.50^{00}-$R2 & $1.55$ & $1.49$ & $2.8$ & $5.0$ & $0.06$ & $0.08$ \\
\hline
MS1b$-e0.60^{00}-$R2 & $0.33$ & $0.29$ & $1.2$ & $2.2$ & $0.10$ & $0.10$ \\
\toprule
MS1b$-e0.45^{\uparrow\uparrow}-$R2 & $1.72$ & $1.69$ & $6.6$ & $11.8$ & $0.10$ & $0.10$ \\
\hline
MS1b$-e0.50^{\uparrow\uparrow}-$R2 & $0.90$ & $0.85$ & $2.0$ & $3.5$ & $0.08$ & $0.09$ \\
\hline
MS1b$-e0.60^{\uparrow\uparrow}-$R2 & $3.49$ & $3.15$ & $12.9$ & $23.1$ & $0.10$ & $0.11$ \\
\hline
\end{tabular}
\end{small}
\end{table}

The ejecta masses computed using the two methods briefly described in 
Sec.~\ref{sec:simu} are given in Table~\ref{tab:ejecta}, 
where $M_\text{ej}^{\mathcal{V}}$ denotes the volume integrated ejecta mass 
and $M_\text{ej}^{\mathcal{S}}$ the ejecta mass computed via Eq.~\eqref{eq:Mejecta_S},
the method of integrating the flux of unbound matter through coordinate spheres. 
We find good agreement between the two ejecta mass estimates,
with differences below 11\%.

Considering the effect of resolution, we find that exact
quantitative statements about the exact ejecta mass cannot be made and the 
discussion should just be seen as qualitative. This is often the case for
computations of ejecta masses using full 3D NR simulations. Even though
simulation methods are continually being improved and simulations are
achieving better and better accuracies, the quantification of 
ejecta material is still challenging and results come with 
large error bars. It is well known in the NR community that the 
accuracy of the NR data for quantities such as the 
unbound mass and kinetic energy 
have uncertainties which range between $\sim$10\% up to even $\sim$100\%, 
see, e.g., Appendix~A of~\cite{Hotokezaka:2012ze,Shibata:2017xdx}
and~\cite{Dietrich:2016fpt,Abbott:2017wuw} for more discussions. Even though 
there are large 
uncertainties in the predictions for the ejecta masses, 
it nevertheless behooves us to understand at least qualitatively 
the dynamical ejecta mechanisms for eccentric binaries. 

We find for the nonspinning SLy (soft EOS) case that the ejecta mass decreases 
as the eccentricity is increased keeping the initial separation fixed.
This is in agreement 
with~\cite{East:2012ww} where equal-mass NSs with comparable compactness 
of 0.17 and mass of 1.35 have been studied. 
Reference~\cite{East:2012ww} showed that for a decreasing impact parameter $r_p$ 
(equivalent to increasing eccentricity for our cases, see Fig.~\ref{fig:SLy:properdistance}) 
the amount of unbound matter decreases. 
For the stiffer EOS (MS1b) setup, 
we find that they have slightly more unbound matter as compared 
to the SLy configurations, which is in agreement with, e.g.,~\cite{Dietrich:2016fpt}
and is caused by larger tidal tail ejecta. 

We find that in the cases where the 
ejecta mass is $\sim \mathcal{O}(10^{-2})M_{\odot}$, 
apart from ejecta from tidal 
tail there is also some ejecta that comes out in the 
merger-postmerger phase either due to shock heating~\cite{Rosswog:1998hy,Oechslin:2006uk,
Bauswein:2013yna,Wanajo:2014wha,Sekiguchi:2015dma,
Hotokezaka:2015cma,Rosswog:2016dhy,Wollaeger:2017ahm,Bovard:2017mvn}
when the cores of the two NS collide or from redistribution of 
the angular momentum within the postmerger remnant; see, e.g.,~\cite{Radice:2018xqa}. 

In comparison with quasicircular binaries, the MS1b cases 
have ejecta masses about one order of magnitude larger, cf.~\cite{Dietrich:2016fpt}. 
Similarly, for the SLy case we find that the ejecta mass is slightly larger for 
most setups as compared to the analogous quasicircular case, but of the same
order [$\mathcal{O}(10^{-2})M_{\odot}$]. 
However, considering the difference among eccentric setups with
fixed initial separation, 
no strong correlation between the exact eccentricity and 
the ejected mass is visible for all the configurations 
except the ones with SLy EOS and no spin. 

Interestingly, another source for small amounts 
of unbound matter is grazing close encounters before the merger 
cf.~the \textit{top-left} and \textit{top-right} panels 
of~Fig.~\ref{fig:2dtracks:SLy00}. 
As the stars undergo more frequent encounters, 
the unbound matter 
increases from $\mathcal{O}(10^{-4})M_{\odot}$ 
to $\mathcal{O}(10^{-3})M_{\odot}$ 
until merger. Overall, the unbound material at the merger 
[$\mathcal{O}(10^{-3})M_{\odot} - \mathcal{O}(10^{-2})M_{\odot}$] is found to 
be ejected as a mildly relativistic and mildly isotropic outflow with the 
velocities $\sim 0.06c-0.15c$. 

\subsection{EM counterparts}
\label{subsec:EM}

As in Refs.~\cite{Dietrich:2016hky,Dietrich:2016lyp}, 
we want to present order-of-magnitude estimates for 
possible electromagnetic counterparts to the merger of
these eccentric binary neutron stars.

\begin{figure}[t]
\includegraphics[width=0.5\textwidth]{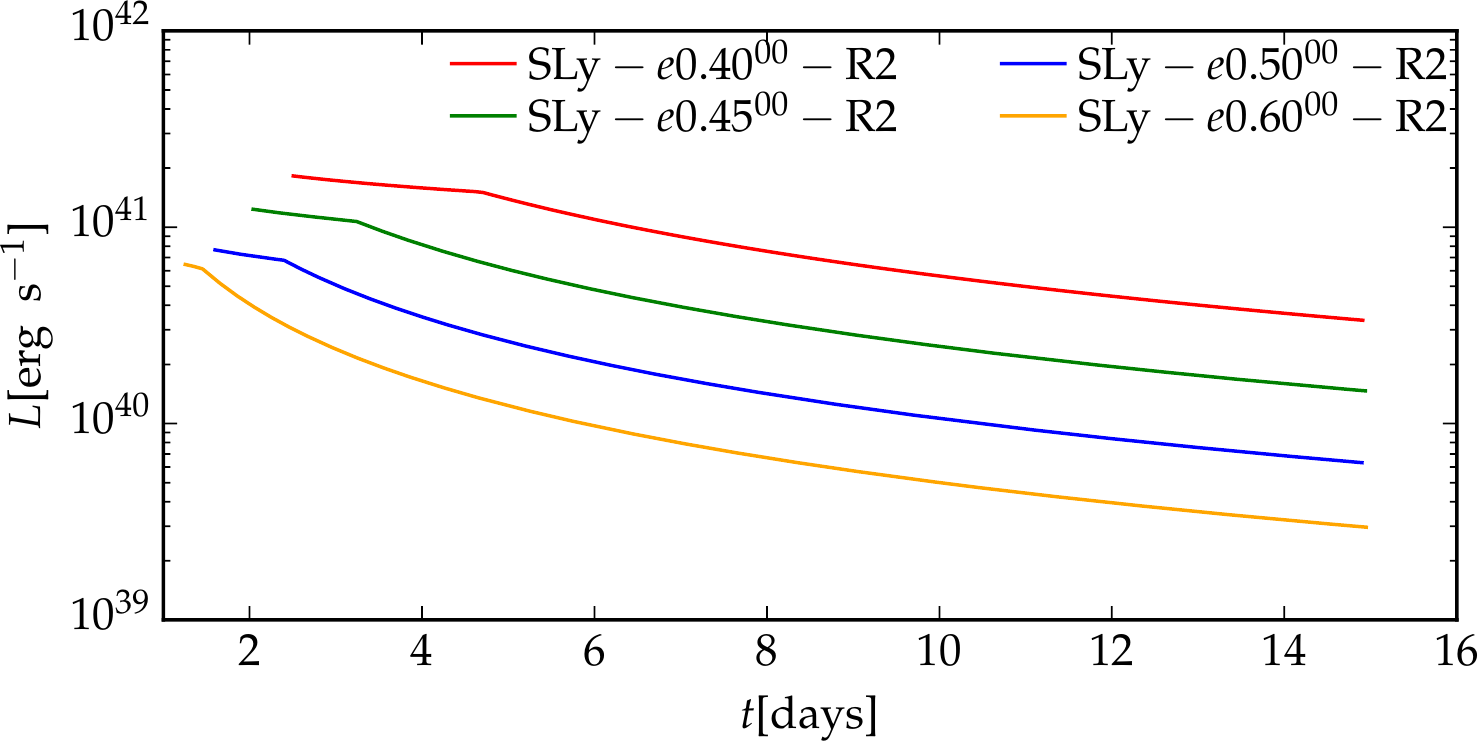}
\caption{
Bolometric luminosity for the SLy setups with zero spin 
configurations and varying initial eccentricity. 
The luminosities are computed using the publicly available BNS 
Kilonova Lightcurve Calculator~\cite{BNSLightCurve}. 
The initial times for the bolometric lightcurves differ 
due to the different ejecta masses; see discussion 
in~\cite{Dietrich:2016fpt}.
}
\label{fig:EM:SLy_lbol}
\end{figure}

\begin{figure}[t]
\includegraphics[width=0.5\textwidth]{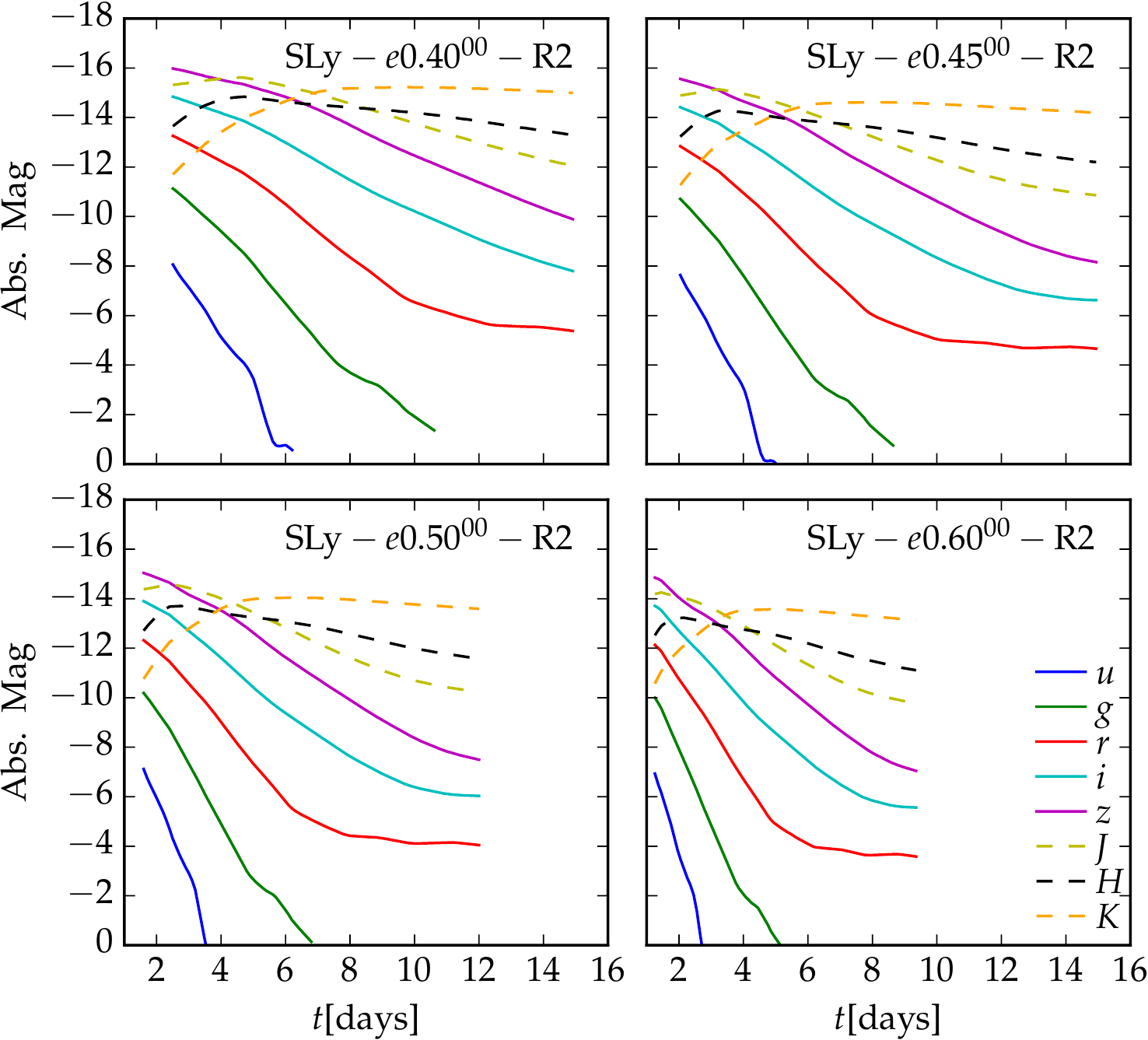}
\caption{
Absolute magnitudes in the ugriz-bands and JHK-bands 
for the SLy setups with zero spin configurations and 
varying initial eccentricity. 
The magnitudes are computed with the BNS Kilonova Lightcurve 
Calculator available at~\cite{BNSLightCurve}; see 
Ref.~\cite{Dietrich:2016fpt} for details.}
\label{fig:EM:SLy_ugrizJHK}
\end{figure}

\paragraph*{\textbf{Kilonovae:-}}

As a starting point, we follow~\cite{Grossman:2013lqa} and present a simple estimate for
the time $t_\text{peak}^\text{NIR}$ at which the peak in the 
near-infrared occurs, as well as an estimate of the corresponding bolometric luminosity 
at this time $L_\text{peak}^\text{NIR}$, and the corresponding temperature $T_\text{peak}^\text{NIR}$.
Table~\ref{tab:EMlightcurve} summarizes our findings. Overall, we find compatible 
results for quasicircular and eccentric BNS systems. 

In addition to the estimates of the peak time, luminosity, and temperature, 
we also want to present simple estimates for the time evolution of the bolometric luminosity 
and the lightcurves in different bands. 
For this purpose we rely on the analytic approximations of~\cite{Dietrich:2016fpt}
and use the publicly available BNS Kilonova Lightcurve Calculator~\cite{BNSLightCurve}.
This approach neglects the composition of the ejecta 
(which is also not evolved in our numerical relativity simulations) 
and focused on the dynamically ejected matter released 
during the merger process. 
Input parameters are taken from Table~\ref{tab:ejecta}. 
The latitudinal and the longitudinal opening angles are 
estimated by evaluation of Eqs.~(12) and~(13) of 
\cite{Dietrich:2016fpt}. 
Furthermore, we use $\kappa = 10 \ {\rm cm}^2 \text{ g}^{-1}$ for the 
opacity, $ \epsilon _0 = 1.58 \times 10^{10} \ {\rm erg} \text{ g}^{-1} \text{ s}^{-1}$ 
for the heating rate coefficient, $\alpha = 1.3$ 
for the heating rate power, and $\epsilon _{th} = 0.5$ 
for the thermalization efficiency as 
in~\cite{Dietrich:2016hky,Dietrich:2016lyp}.
Figure~\ref{fig:EM:SLy_lbol} and Fig.~\ref{fig:EM:SLy_ugrizJHK} 
present our results for the time evolution of the bolometric luminosities and the absolute 
magnitudes for the ugrizJHK bands~\cite{Fukugita:1996qt}.
We find that the configurations we consider will have a luminosity between $10^{39}-10^{42}$ 
erg s$^{-1}$ over a time ranging from a few days to two weeks after the merger (Fig.~\ref{fig:EM:SLy_lbol}).
This will in general be true for all the configurations as the 
luminosity strongly correlates with the mass of the ejecta, which we have already seen to
be~$\mathcal{O}(10^{-3})-\mathcal{O}(10^{-2})M_{\odot}$. 
Therefore, we find in our setups that for an increasing 
eccentricity the luminosity decreases by more 
than one order of magnitude for our nonspinning
configurations employing the SLy EOS.
For other configurations there is no strong correlation 
between the ejecta masses and the initial 
eccentricities, as discussed in 
Sec.~\ref{subsec:ejecta}.

\paragraph*{\textbf{Radio flares:-}}
In order to estimate the radio emission caused by the mildly relativistic
ejecta, we use the model of~\cite{Nakar:2011cw} 
which uses as input variables from our simulations the kinetic 
energy and the velocity of the ejecta. We summarize the
quantities for all the configurations studied
in this work in Table~\ref{tab:EMlightcurve}. Most notably
we find that radio flares will have largest fluence at
$t^\text{radio} _\text{peak}$ $\sim\mathcal{O}(\rm {years})$ 
similar to the quasicircular case~\cite{Dietrich:2016hky,Dietrich:2016lyp}.

\begin{table}[tp]
\caption{
Properties of electromagnetic counterparts. The columns refer to: 
the name of the configuration, the time in which the peak 
in the near infrared occurs $t_\text{peak}^\text{NIR}$, the corresponding 
peak luminosity $L_\text{peak}^\text{NIR}$, the temperature at this time $T_\text{peak}^\text{NIR}$, 
the time of peak in the radio band $t^\text{radio} _\text{peak}$, 
and the corresponding radio fluence (the flux density per unit frequency)
$F^{\nu \text{radio}} _\text{peak}$ at 100 Mpc. 
We present results for all resolutions.}
\label{tab:EMlightcurve} 
\centering
\begin{tabular}{cccccc}
\toprule
Name & $t_\text{peak}^\text{NIR}$ & $L_\text{peak}^\text{NIR}$ & $T^\text{NIR}_\text{peak}$ & $t^\text{radio} _\text{peak}$ & $F^{\nu \text{radio}} _\text{peak}$\\
$ $ & $[\rm {days}]$ & $[\rm {10^{40}erg \text{ s}^{-1}}]$ & $[\rm {10^3 K}]$ & $[\rm {years}]$ & $[\rm {mJy}]$\\
\hline 
SLy$-e0.40^{00}-$R2 & $5.4$ & $4.5$ & $1.9$ & $15.7$ & $0.860$\\
\hline
SLy$-e0.45^{00}-$R1 & $5.4$ & $4.4$ & $1.9$ & $10.2$ & $0.208$\\

SLy$-e0.45^{00}-$R2 & $3.8$ & $3.7$ & $2.0$ & $6.8$ & $0.097$\\

SLy$-e0.45^{00}-$R3 & $4.9$ & $5.0$ & $1.9$ & $6.9$ & $0.222$\\

SLy$-e0.45^{00}-$R4 & $3.6$ & $4.3$ & $2.0$ & $6.1$ & $0.251$\\
\hline
SLy$-e0.50^{00}-$R1 & $5.0$ & $2.2$ & $2.3$ & $12.9$ & $0.003$\\

SLy$-e0.50^{00}-$R2 & $2.8$ & $2.5$ & $2.4$ & $6.0$ & $0.019$\\

SLy$-e0.50^{00}-$R3 & $6.0$ & $3.7$ & $1.9$ & $18.1$ & $0.162$\\

SLy$-e0.50^{00}-$R4 & $2.1$ & $2.4$ & $2.5$ & $4.7$ & $0.024$\\
\hline
SLy$-e0.60^{00}-$R2 & $1.7$ & $2.2$ & $2.7$ & $3.9$ & $0.025$\\
\toprule
SLy$-e0.40^{\uparrow \uparrow}-$R2 & $4.5$ & $4.4$ & $1.9$ & $7.5$ & $0.184$\\
\hline
SLy$-e0.45^{\uparrow \uparrow}-$R1 & $4.0$ & $3.1$ & $2.2$ & $8.1$ & $0.032$\\

SLy$-e0.45^{\uparrow \uparrow}-$R2 & $2.2$ & $2.1$ & $2.6$ & $5.5$ & $0.014$\\

SLy$-e0.45^{\uparrow \uparrow}-$R3 & $3.1$ & $3.7$ & $2.1$ & $5.6$ & $0.156$\\

SLy$-e0.45^{\uparrow \uparrow}-$R4 & $3.3$ & $2.8$ & $2.3$ & $7.4$ & $0.034$\\
\hline
SLy$-e0.50^{\uparrow \uparrow}-$R2 & $3.7$ & $2.5$ & $2.3$ & $8.8$ & $0.012$\\
\hline
SLy$-e0.60^{\uparrow \uparrow}-$R2 & $4.1$ & $3.6$ & $2.1$ & $7.3$ & $0.063$\\
\toprule
MS1b$-e0.45^{00}-$R2 & $5.2$ & $2.6$ & $2.2$ & $13.2$ & $0.011$\\
\hline
MS1b$-e0.50^{00}-$R2 & $6.1$ & $2.9$ & $2.1$ & $14.0$ & $0.013$\\
\hline
MS1b$-e0.60^{00}-$R2 & $2.4$ & $2.1$ & $2.6$ & $6.0$ & $0.013$\\
\toprule
MS1b$-e0.45^{\uparrow \uparrow}-$R2 & $5.4$ & $3.8$ & $2.0$ & $10.5$ & $0.068$\\
\hline
MS1b$-e0.50^{\uparrow \uparrow}-$R2 & $4.2$ & $2.7$ & $2.2$ & $9.2$ & $0.014$\\
\hline
MS1b$-e0.60^{\uparrow \uparrow}-$R2 & $7.5$ & $5.0$ & $1.7$ & $12.1$ & $0.149$\\
\hline
\end{tabular}
\end{table}

\section{Gravitational Waves}
\label{sec:GWs}

\subsection{Inspiral}

\begin{figure*}[t]
\includegraphics[width=\textwidth]{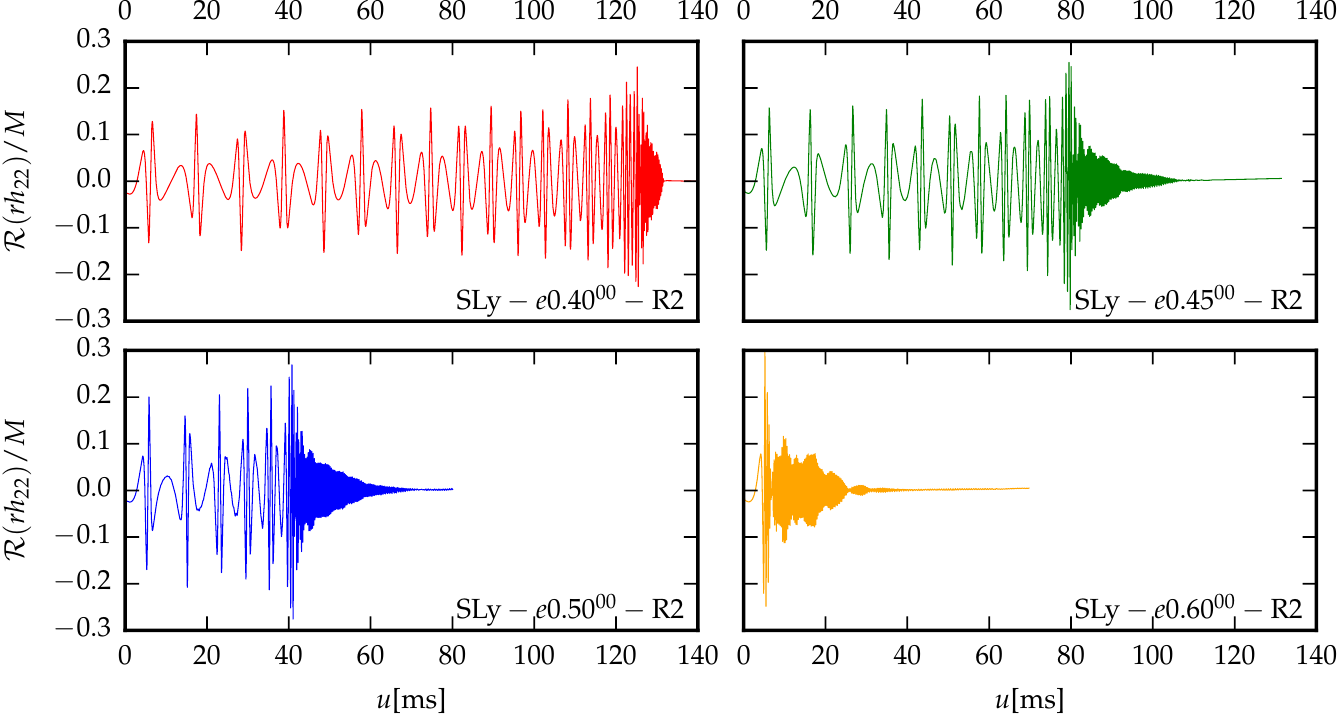}
\caption{
Real part of the (2,2) mode of the gravitational 
wave strain $rh$ vs.\ the retarded time $u$. The color code refers to Fig.~\ref{fig:tracks:SLy00} 
and corresponds to different eccentricities. 
See, e.g., \cite{Hinder:2017sxy} for eccentric BBH waveforms, for comparison (albeit for smaller
eccentricities).}
\label{fig:Reh_SLy00}
\end{figure*}

We extract GWs and metric multipoles following Ref.~\cite{Brugmann:2008zz}.
The $\l=m=2$ multipoles of the GWs extracted at $r=900M_{\odot}$ are shown in 
Fig.~\ref{fig:Reh_SLy00} for the configurations employing SLy EOS and no spin. 
The metric multipoles $rh_{\lm}$ are reconstructed from the curvature multipoles 
using the fixed frequency integration of \cite{Reisswig:2010di}. 
We set the low-frequency cutoff to be half the initial GW frequency. 

\begin{figure}[t]
\includegraphics[width=0.5\textwidth]{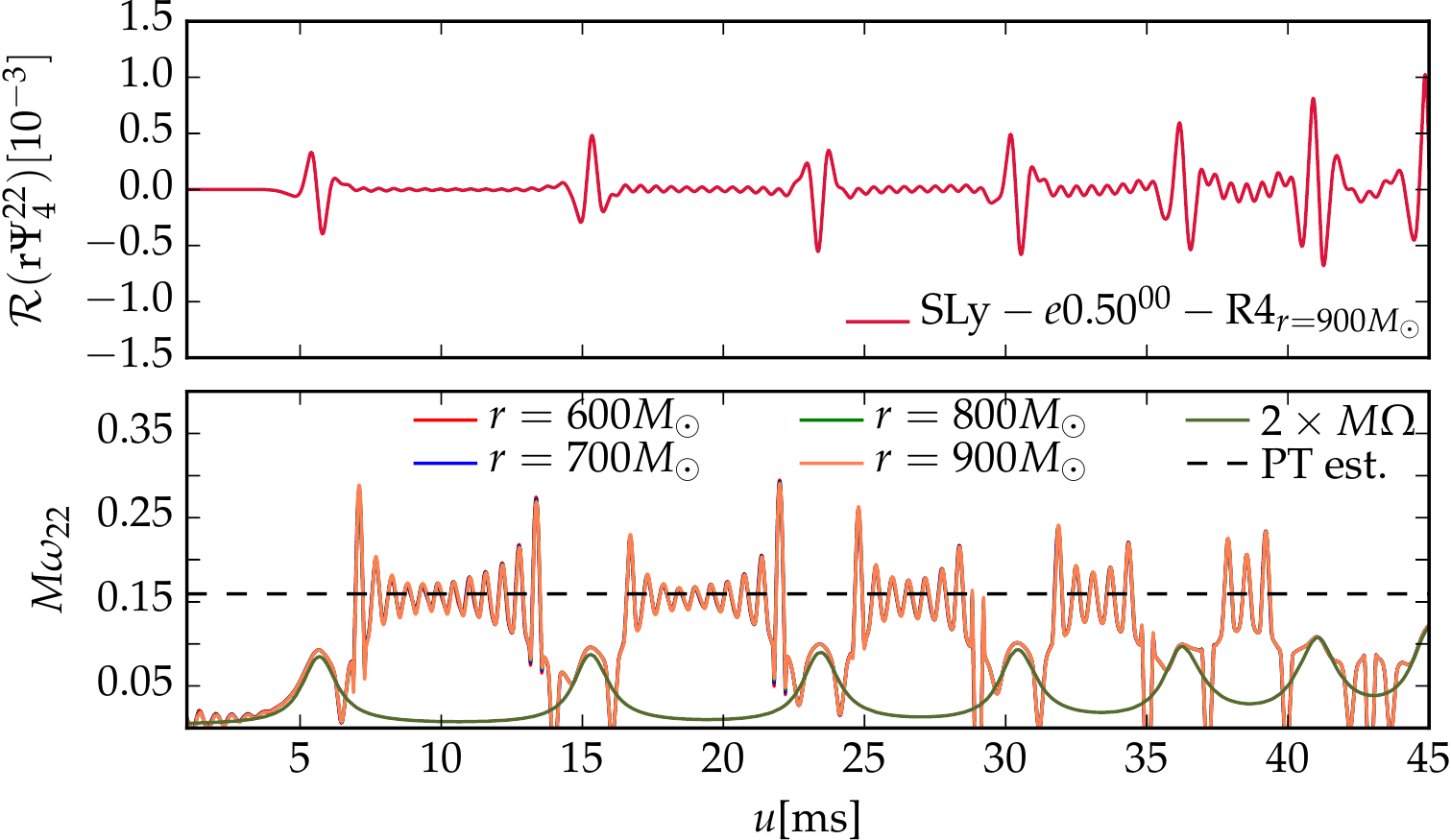}
\caption{
$r\Psi ^{22} _4$ extracted at $r = 900M_\odot$ and instantaneous GW frequency $M\omega_{22}$ computed from
the phase of $r\Psi ^{22} _4$
(note that the curves at different extraction radii are coincident)
plotted alongside $ 2 \times M\Omega$ as computed from the coordinate tracks of the NSs, both for the inspiral part of
the SLy$-e0.50^{00}-{\rm R4}$ case. We find a good agreement for the relation
$M\omega_{22} = 2 \times M\Omega$ and similar results hold for all the
other cases. The higher frequency in-between the periastron encounters
correspond to the quasinormal
mode excitations of the NSs. We also plot the
perturbation theory (PT) estimate for the $f$-mode frequency as a black dashed line.}
\label{fig:Momega22:SLy00}
\end{figure}

The emission from the orbital motion is very different from the 
characteristic chirping signal of quasicircular orbits,
in which frequency and amplitude monotonically increase. 
One of the interesting features is the effect of eccentricity. In Fig.~\ref{fig:Reh_SLy00}, 
one can see that the usual oscillations in the strain at twice the orbital frequency are 
modulated by an oscillating envelope with a frequency lower than the orbital 
frequency, corresponding to apsidal precession of the pericenter. 
A notable feature for BNSs on eccentric orbits is the quasinormal mode oscillations 
of the NSs as already discussed in Sec.~\ref{ssec:sim_analysis}. 
These oscillations are superposed on the GW signal from the binary's orbital
motion. While the NS oscillations are hardly visible in
the metric multipole $rh_{22}$, they are evident in the curvature multipoles. 
In Fig.~\ref{fig:Momega22:SLy00}, we plot the (2,2) mode of $r\Psi _4$
and the corresponding instantaneous GW frequency
for the  SLy$-e0.50^{00}-{\rm R4}$ case, focusing on the inspiral part.
The figure also shows consistency of the extracted GW signals at different extraction radii,
and in general other quantities extracted at finite radii from the simulated BNS system.
In the lower panel of Fig.~\ref{fig:Momega22:SLy00}, we find that
the influence of finite radius extraction of the GWs is negligible.
Therefore, no radius extrapolation to compensate for the finite radius 
extraction~\cite{Bernuzzi:2016pie} is employed.
It is reassuring that the GW frequency computed from the phase 
of the GWs matches twice the orbital frequency computed from 
the star trajectories during close encounters.
Furthermore, comparing with the plot of the instantaneous GW frequency
for eccentric BBHs in Ref.~\cite{Hinder:2017sxy}, 
we find for the BBH case there is no frequency higher than 
twice the orbital frequency whereas
for the BNS case we find a much higher frequency due to the $f$-mode oscillations
of the stars.

%\blue{all \textit{f}-modes have frequencies lower than 4 kHz~\cite{Kokkotas:1999mn}}.

\subsection{\textit{f}-mode oscillations}

\begin{figure*}[t]
\includegraphics[width=\textwidth]{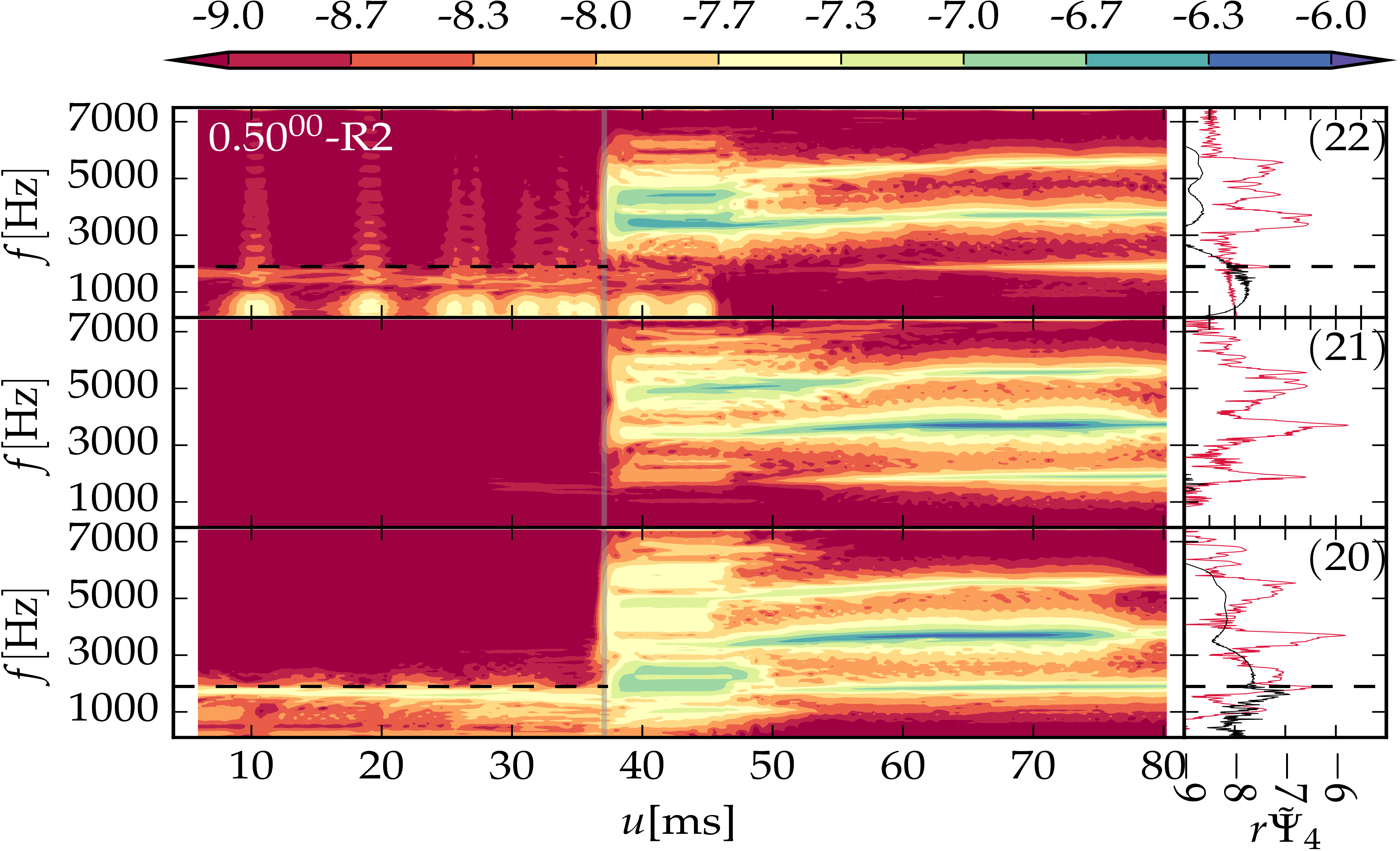}
\caption{
Spectrogram (left) and power spectral density (right) of the curvature 
scalar $r\Psi _4$ for the SLy$-e0.50^{00}-{\rm R2}$ configuration.
The labels in the right panel plot refer to the (22), (21), and (20) modes
of the curvature scalar. The black dashed line marks 
the PT estimate of the \textit{f}-mode frequency
and the gray line marks the moment of merger. 
The PSD is split into a part before the moment of merger (black) and a part 
after the merger (crimson).}
\label{fig:spectrogram:SLy00}
\end{figure*}

\begin{figure*}[t]
\includegraphics[width=\textwidth]{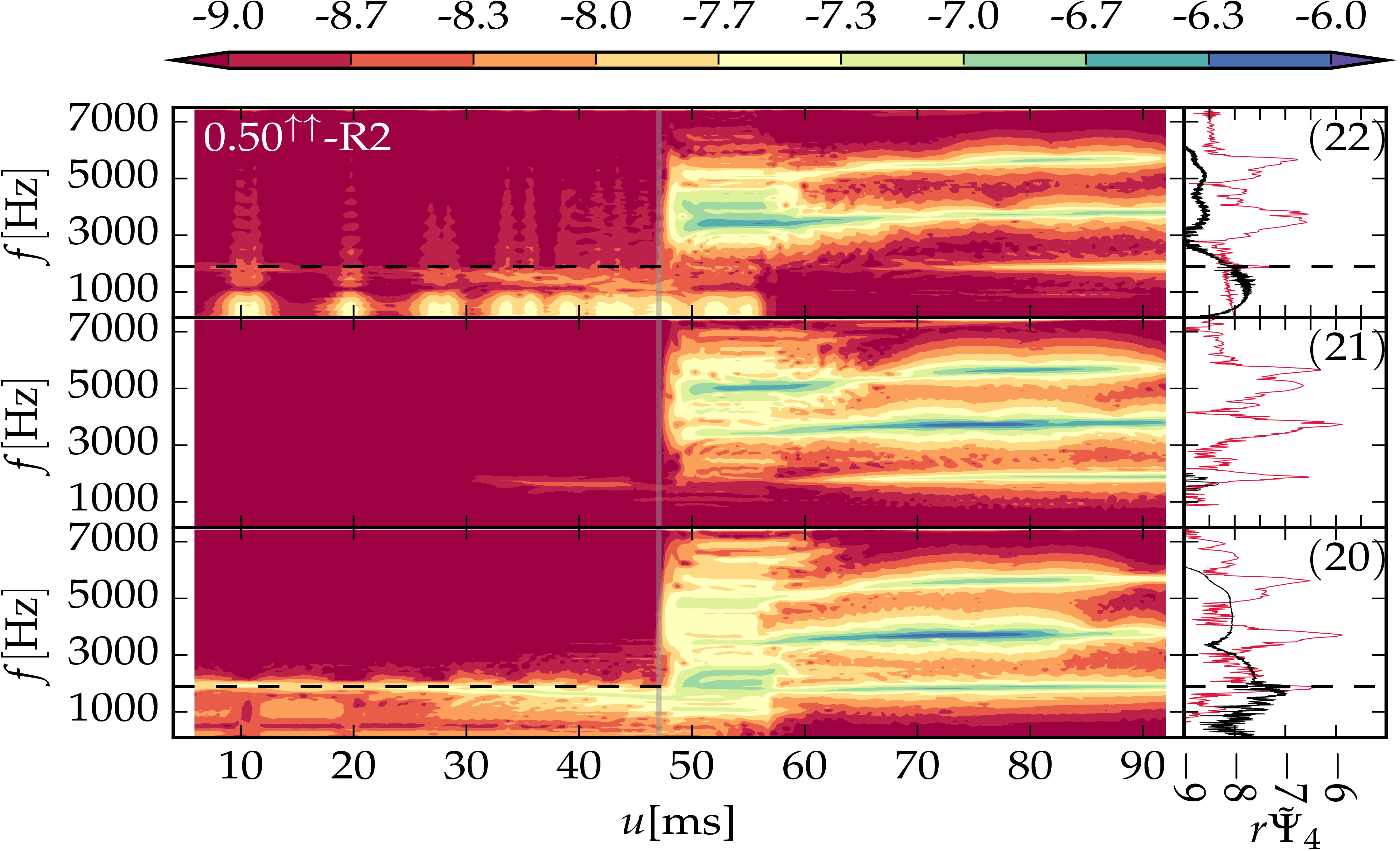}
\caption{
Spectrogram (left) and power spectral density (right) of the curvature 
scalar $r\Psi _4$ for the SLy$-e0.50^{\uparrow \uparrow}-$R2 configuration.
The labels in the right panel plot refer to the (22), (21), and (20) modes
of the curvature scalar.
The black dashed line marks the 
PT estimate of the \textit{f}-mode frequency for \emph{nonspinning} stars
and the gray line marks the moment of merger. The PSD is split into a part 
before the moment of merger (black) and a part after the merger (crimson).}
\label{fig:spectrogram:SLyuu}
\end{figure*}

Figure~\ref{fig:spectrogram:SLy00} and Fig.~\ref{fig:spectrogram:SLyuu} 
(left panels) show the (normalized) spectrogram for
SLy$-e0.50^{00}-{\rm R2}$ and SLy$-e0.50^{\uparrow \uparrow}-{\rm R2}$,
respectively. The right panels show the PSDs 
for the premerger (black) and postmerger 
(crimson) phases as different colors for the 
individual modes. 
The color bar in the spectrogram goes 
from red to blue and is given in arbitrary 
units, since we are only interested in 
the frequencies and the relative strength. 

During the inspiral one observes discrete GW bursts present in the $(2,2)$
mode. This is in contrast to the typical chirp signal for the quasicircular
orbits, where the frequency and the amplitude increase monotonically over
time. We also find a low power, but higher frequency region (1.5~kHz - 2~kHz)
distinct from the inspiral burst signals. Such a frequency region is prominent
for cases where the NSs' distance decreases to values as small as 40km-60km (cf.\
Fig.~\ref{fig:SLy:properdistance}) during the periastron encounters. Thus,
we are easily able to extract the NS oscillation frequencies for the
configurations for which we set the initial eccentricity, $e$, to be $0.45$ or
$0.50$.

The matter mode excitations can be reliably confirmed by studying the
spectra of the $(2,0)$ fluid mode excitations. For BNS (and BHNS)
in eccentric orbits, the $(2,0)$ fluid
mode is expected to be excited as the stars undergo
periastron passage (see, e.g., the
Newtonian calculations in~\cite{Yang:2018bzx}). However,
the spectrogram in the bottom panel of
Fig.~\ref{fig:spectrogram:SLy00} for the $(2,0)$
curvature scalar mode, where a relatively high power signal is
visible at $\sim 1.8$kHz during the inspiral phase is accounted
almost entirely by mode-mixing from (2,$\pm$2) modes, as we will see in
the following discussion.\footnote{Note that for the current
set of simulations we do not have the 3D data which is
required for studying the fluid mode oscillations and
therefore we delay such an analysis to future study.}
We compute the perturbation theory (PT) estimate of the $f$-mode 
excitation frequency for the SLy EOS with no spin of the NSs using the \textit{f}-Love relation
from~\cite{chan:2014kua},\footnote{We use
the publicly available TOV solver of Ref.~\cite{Bernuzzi:2014owa,TEOBresum}
to compute the Love number.} obtaining an \textit{f}-mode frequency of $1.89$~kHz. As visible in the
spectrogram, Fig.~\ref{fig:spectrogram:SLy00}, there is good agreement between the NS oscillation frequency from the simulation 
and the PT estimate of the \textit{f}-mode frequency (horizontal black dashed line). Some of the differences between the PT estimate and the spectrogram are
attributable to the gravitational redshift due to the star's companion, as discussed below, while others are due to the relatively low resolution of the simulation (R2): 
The \textit{f}-mode frequency observed in the simulations
increases with resolution.

In the spinning case, we find in Fig.~\ref{fig:spectrogram:SLyuu} that the bright area near $1.8$kHz in the $(2,2)$ and the $(2,0)$ mode in
Fig.~\ref{fig:spectrogram:SLy00} for the nonspinning case is shifted to
slightly higher frequencies. Notice that the dashed line in both figures
shows the PT estimate for the nonspinning configuration
for easier comparison. Overall, we find similar results for the 
simulations performed at different resolutions 
for the same configurations, see Appendix~\ref{app:conv}.

To obtain a PT estimate for the \textit{f}-mode oscillation frequency 
of spinning configurations, we follow Doneva~\emph{et al.}~\cite{Doneva:2013zqa}. Since we find an increase in the
\textit{f}-mode frequency with spin, we want to consider the $\ell = 2$, $m = -2$ mode\footnote{Since the stellar modes
describe real quantities, e.g., the perturbations to the star's density, the $m = -2$ mode of the star will have an angular dependence of $e^{\pm 2i \phi}$, and
thus its radiation will have a significant overlap with the $(2,2)$ spin-$(-2)$-weighted spherical harmonic mode.}. Additionally, it is expected that this mode will be excited most strongly in highly eccentric binaries, as discussed 
in~\cite{Yang:2018bzx}, using Newtonian calculations.
For our case, the NSs with SLy EOS are spinning at $f_\star \simeq 191$Hz. Since the Doneva~\emph{et al.}\ results relate the \textit{f}-mode frequency
of a spinning star to that of a nonspinning star with the same central density, we first note that a nonspinning SLy star with the same central density as the spinning star has a
mass of $1.42M_\odot$ and thus a \textit{f}-mode frequency of $f^\text{Cowling}_\text{nonspinning} = 1.93$~kHz (obtained using the \textit{f}-Love relation from~\cite{chan:2014kua}). We then use Eq.~(24) in Doneva~\emph{et al.}~\cite{Doneva:2013zqa}, which gives
\begin{equation}
\frac{\omega^\text{corot, Cowling}_\text{spinning}}{\omega^\text{Cowling}_\text{nonspinning}} = 1 - 0.235\frac{\Omega_\star}{\Omega _\text{K}} - 0.543\left(\frac{\Omega_\star}{\Omega _\text{K}}\right)^2.
\end{equation}
Here $\omega^\text{corot, Cowling}_\text{spinning} = \omega^\text{Cowling}_\text{spinning} - 2\Omega_\star$ is the mode's angular frequency in the frame corotating with the star, where $\omega^\text{Cowling}_\text{spinning}$ is the mode's angular frequency in the inertial frame of an external observer and $\Omega_\star = 2\pi f_\star$ is the angular velocity of the star. We have $\Omega_\star/ \Omega _\text{K} = 0.157$ for these SLy stars, where $\Omega _\text{K}$ is the Kepler angular velocity for an SLy star with the same central density as the stars we consider (computed using LORENE~\cite{LoreneCode}). Additionally, we have used superscripts of ``Cowling'' to denote that the expression in Doneva~\emph{et al.}\ is derived using the Cowling approximation.

The Cowling approximation generally overestimates the \textit{f}-mode frequency, as illustrated in, e.g., Fig.~5 in Ref.~\cite{Zink:2010bq}. However, this figure shows that this overestimate is independent of spin (to a good approximation, particularly for the relatively small spins we are considering). Thus, we can use the Cowling approximation offset of $\Delta f_\text{Cowling} \simeq 500$~Hz for a $1.42M_\odot$ nonspinning star obtained from Fig.~8 in Ref.~\cite{Chirenti:2015dda} to correct for the effect of the approximation (which is, however, only a $\sim 1\%$ effect on the final value). Specifically, if we write $\omega^\text{corot, Cowling}_\text{spinning} =: k\omega^\text{Cowling}_\text{nonspinning}$ (so $k = 0.95$ here), we have
\begin{equation}
\omega_\text{spinning} ^{f\text{-mode}} = k \omega_\text{nonspinning} ^{f\text{-mode}} + 2\pi (k-1)\Delta f_\text{Cowling} +2\Omega_\star.
\end{equation}
This procedure gives a \textit{f}-mode frequency of $2.20$~kHz.

The redshifted PT estimate of the frequency is $\sim 5\%$ larger than the frequency observed in the spectrogram or the instantaneous frequency we compute using the method given below. One would need the spin of the stars to
be $\sim 40\%$ smaller than its actual value in order for the instantaneous frequency estimated from the waveform to agree with the redshifted PT frequency. Such a large
difference in spin is well outside the maximum expected difference between the true value of the spin and the one estimated from the inputs to the initial data construction, as discussed in Ref.~\cite{Dietrich:2016lyp}.
(Note that the spin we estimate from the initial data inputs agrees well with the one we compute in the early part of the inspiral using the quasilocal computation described in Ref.~\cite{Dietrich:2016lyp}.)
Thus, we are not sure of the source of the discrepancy between the PT estimate and the frequency of the oscillations observed 
in the simulation.\footnote{A potential source of the disagreement might be the 
change in the external gravitomagnetic fields of the NSs due to 
the intrinsic NS spins~\cite{Steinhoff:2016rfi}. 
However, this effect does not seem large enough to account for the observed discrepancy.} 
Additionally, there do not appear to be any other $m = \pm 2$ modes
that would occur at the observed frequency.

\begin{figure}[t]
\includegraphics[width=0.475\textwidth]{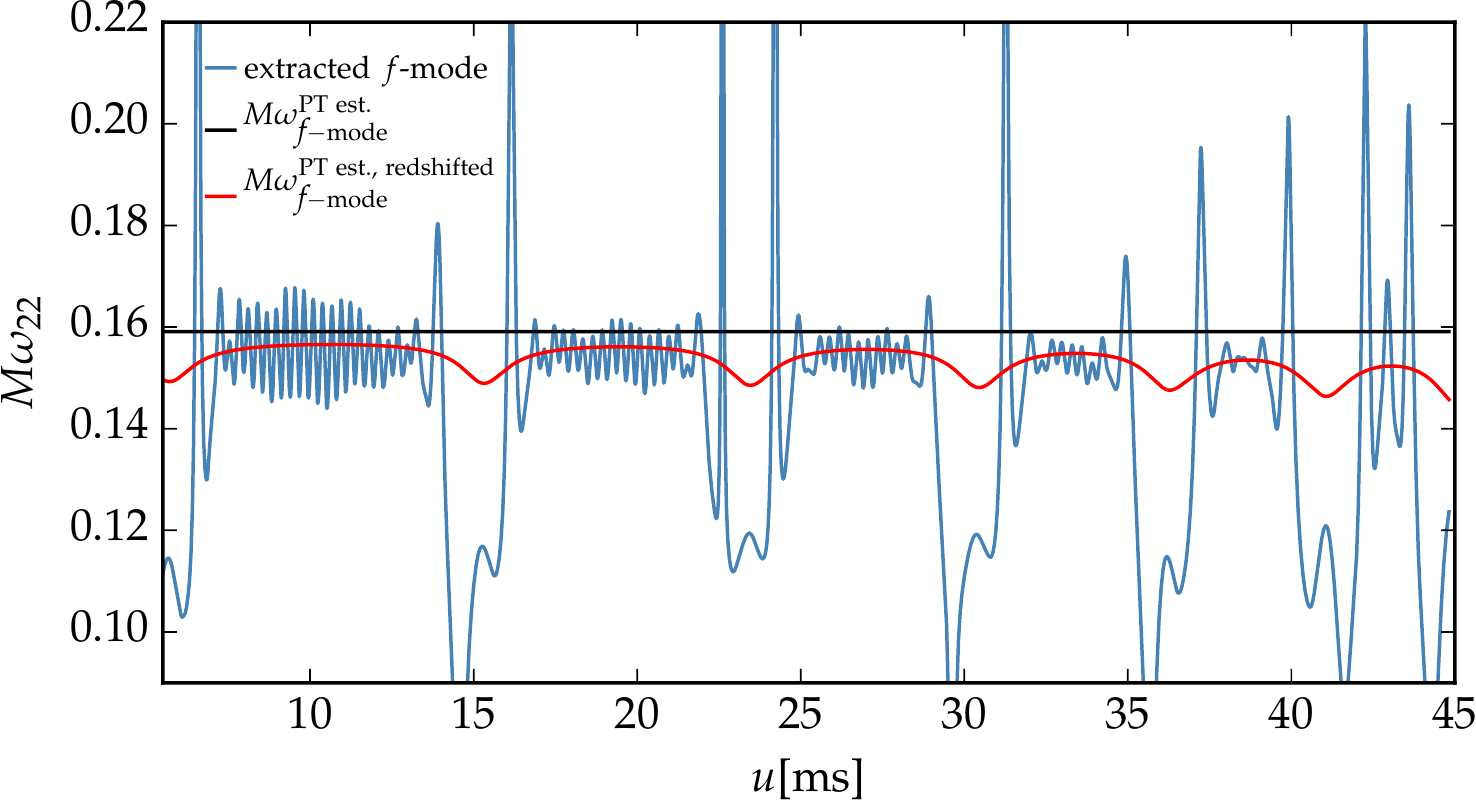}
\caption{
Instantaneous frequency of $\Psi^{4, f\text{-mode}}_{2,2}$ 
for SLy$-e0.50^{00}-$R4, compared to the prediction of 
the \textit{f}-mode frequency of an isolated star, as well as the
\textit{f}-mode frequency including the estimated gravitational redshift due to the
star's companion. These curves should only be compared away from
the periastron bursts.
}
\label{fig:Momega22_illustration}
\end{figure}

In order to see the effects of the redshift, we can examine the instantaneous \textit{f}-mode frequency we obtain 
from the $(2,2)$ mode after removing the orbital contribution (as described in Sec.~\ref{ssec:sim_analysis}). This is shown in 
Fig.~\ref{fig:Momega22_illustration}. (Since $\mu_{2,2;2,2}$ is real, 
accounting for mode mixing does not change the instantaneous frequency here.) 
Here we estimate this redshift using 
the stars' tracks and the leading PN expression of 
$\omega_{f\text{-mode}}^\text{redshifted} = (1 - M^B/d)\omega_{f\text{-mode}}$
(see, e.g., Ref.~\cite{1993PhRvD..48.4639K}), where $d$ is the separation of the two stars,
noting that it suffices to consider only star~$A$, as the binary is symmetric.\footnote{Note that Ref.~\cite{1993PhRvD..48.4639K} calculates higher
PN corrections, through $O(c^{-4})$, including effects of the star's velocity. [See Eq.~(4.1), noting
that $\beta = \gamma = 1$ and the other parameterized PN parameters vanish in general relativity.] The
contributions from the velocity are all negligible
here (producing almost indistinguishable curves on this plot), which is why we do not include them.
The leading $O(v)$ effect from the star's velocity only affects the phase of the $(2,2)$ mode of the
waveform on the timescale of the orbit, and thus does not affect the \textit{f}-mode signal we
consider here. The star's velocity is small enough that the $O(v^2)$ terms produce differences
of $\sim 1\%$. We do not consider the additional corrections involving the gravitational potential, as they are
expected to be small in the region between periastra. Moreover, it is unclear if adding higher corrections
would improve the accuracy of the predictions, as we are not evaluating the expression using PN coordinates.}
The remaining oscillations of the
instantaneous frequency are likely due to a combination of the effects mentioned below in the 
discussion of the $(2,0)$ mode amplitude, as well as the lack of removal of 
the $(2,2) \leftrightarrow (2,-2)$ mixing (which does not seem straightforward to remove), 
and possibly also mixing in of intrinsic higher-$\ell$ modes.\\

\paragraph*{\textbf{Removing displacement-induced mode mixing in the GW signal from tidally-induced
oscillations:-}}

We now want to apply the displacement-induced mode mixing analysis from Sec.~\ref{ssec:sim_analysis} to obtain the
dominant $(2,\pm2)$ modes of the $\textit{f}$-mode oscillations that would be extracted if the stars
were at rest at the origin. We will then use the amplitude of these modes to estimate the energy stored
in the $\textit{f}$-mode oscillations in the next subsubsection.

If we just have a $(2,\pm 2)$ intrinsic excitation of the stars, we will also obtain a purely real contribution to the
$(2,0)$ mode from this intrinsic excitation due to the mode mixing. This contribution is purely real because the
extracted $(2,\pm 2)$ modes have the usual relation for nonprecessing binaries of $\Psi^4_{2,-2} =
\left(\Psi^4_{2,2}\right)^*$ and we have $\mu_{2,-2;2,-2} = \mu_{2,2;2,2}$ and $\mu_{2, -2;2,0} = \mu_{2, 2;2,0}^*$,
so the contribution to the $(2,0)$ mode from mode mixing is
\begin{equation}
\begin{split}
\Psi^{4, \text{mixing from } 2,\pm2}_{2,0} &= \Psi^{4, f\text{-mode}}_{2,2}\mu_{2, 2;2,0}/\mu_{2,2;2,2}\\
&\quad + \Psi^{4, f\text{-mode}}_{2,-2}\mu_{2, -2;2,0}/\mu_{2,-2;2,-2}\\
&= 2\mathcal{R}(\Psi^{4, f\text{-mode}}_{2,2}\mu_{2, 2;2,0}/\mu_{2,2;2,2}).
\end{split}
\end{equation}

\begin{figure*}[t]
\includegraphics[width=\textwidth]{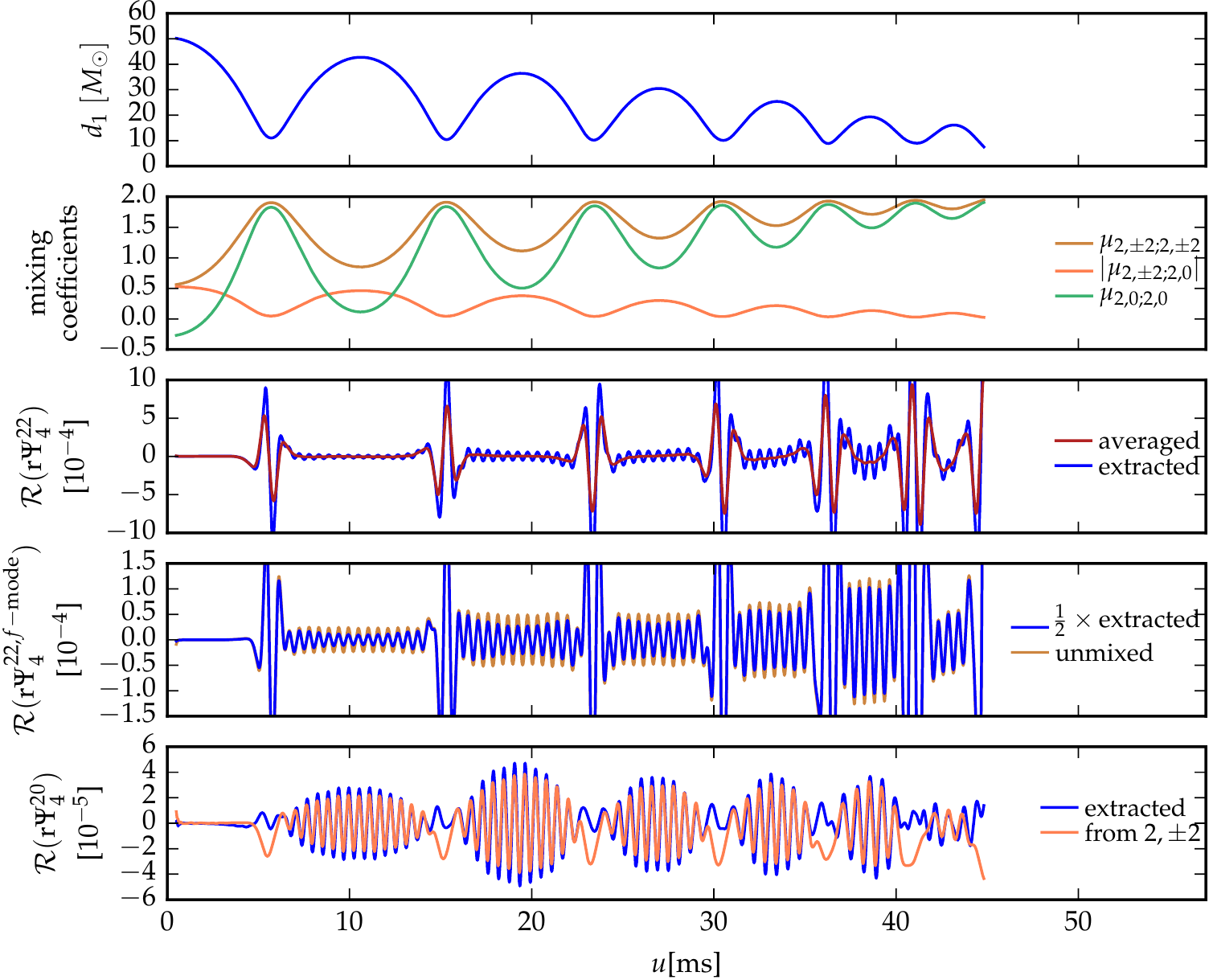}
\caption{
Illustration of the effects of displacement-induced mode mixing on the waveforms extracted from SLy$-e0.50^{00}-$R4: Coordinate distance from star 1 to the origin \emph{(top)}; mode mixing coefficients \emph{(second-from-top)} ($|\mu_{2,\pm 2;2,\mp2}|$ is almost indistinguishable from $0$ at this scale, so we do not show it); illustrating the averaging used to separate the $f$-mode and orbital signals \emph{(third-from-top)}; comparing the full mode mixing calculation of the intrinsic $\textit{f}$-mode contribution to the real part of the $(2,2)$ mode from a single star to the naive calculation of taking half of the extracted mode \emph{(second-from-bottom)} (the result for the imaginary part is exactly analogous); comparing the contribution to the real part of the $(2,0)$ mode from mode mixing to that extracted from the evolution of the binary \emph{(bottom)} (the imaginary part vanishes). In all cases, the approximations used to compute the mode mixing are only valid during the inter-burst $\textit{f}$-mode oscillations, and the predictions of the mode mixing calculation for the periastron bursts should not be considered.}
\label{fig:mode_mixing_illustration}
\end{figure*}

In fact, we find that if we compute the mixing coefficients using 
the tracks to give the positions of the stars, 
the mixing from the $(2,\pm 2)$ modes appears to account for 
all of the \textit{f}-mode signal we extract in the $(2,0)$ mode, 
as illustrated in Fig.~\ref{fig:mode_mixing_illustration}.
[We use the PT computation of the \textit{f}-mode frequency used in
the previous section. We also checked that we find the expected contributions to the $(3,\pm 2)$
and $(3,0)$ modes due to displacement-induced mode mixing from the $(2,\pm 2)$ modes.] 
The slight deviations in amplitude are likely due to the approximation 
of using the coordinate tracks to compute the mixing coefficients
and residual contributions from the orbital motion that are not removed 
by our simple moving average procedure to separate the orbital and \textit{f}-mode 
signals. Additionally, while we expect a contribution to the $(2,0)$ mode from the binary's orbital motion,
we find that any such contribution is considerably smaller than the \textit{f}-mode signal that arises from
displacement-induced mode mixing.

We can also use the Schwarzschild tortoise coordinate computed 
from the tracks instead of the tracks themselves to compute the retarded time, 
using the system's ADM mass as the Schwarzschild mass. 
This is analogous to the procedure used for the extraction of gravitational waves, 
as described in Sec.~V of~\cite{Dietrich:2016hky}, 
though it is less well-motivated here in the stronger-field regime, 
and we simply consider it to give a comparison for 
the results computed using the tracks themselves. 
If we use the tortoise coordinate, then we obtain closer agreement in 
the amplitude of the mixed contribution to the $(2,0)$ mode and 
the extracted contribution, 
but we also obtain intrinsic $(2,\pm 2)$ $f$-mode signals that are 
considerably larger and look rather unphysical, 
since their amplitude increases towards apastron. 
We thus chose to present the results with the plain coordinate track computation.\\

\paragraph*{\textbf{Energy estimate of the NS oscillations:-}}
In order to give an order of magnitude 
estimate of the energy stored in the 
NS oscillations, we assume that it decays exponentially due to the emission
of GWs. We then compute the \textit{f}-mode GW damping time and use it
to infer the energy stored in the NS oscillations by computing the
energy radiated in GWs.
In particular, we compute the $f$-mode angular frequency
using the same \textit{f}-Love relation from Eq.~(3.5) in~\cite{chan:2014kua}
used previously, which gives an \textit{f}-mode frequency of $1.40$~kHz for the nonspinning
MS1b stars. The damping time $\tau_{f\text{-mode}}$ (the inverse of the imaginary part
of the mode's angular frequency) is computed using Eq.~(20) 
in~\cite{Lioutas:2017xtn}, which gives it in terms of the star's mass and radius. We obtain damping times of
$0.186$~s and $0.317$~s, respectively, for the nonspinning SLy and MS1b stars.

\begin{figure*}[t]
\includegraphics[width=\textwidth]{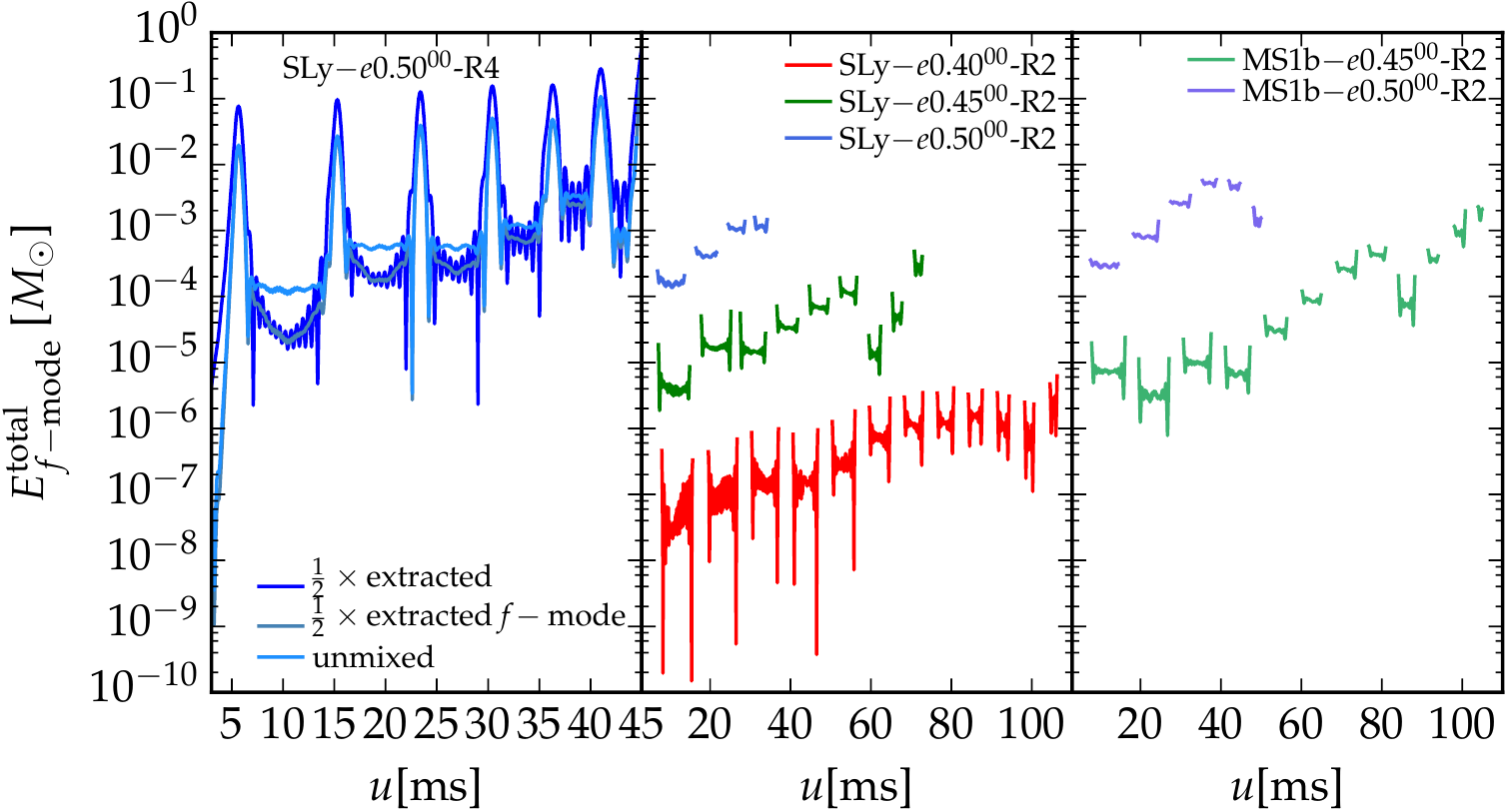}
\caption{
Estimate of the energy stored in \textit{f}-mode oscillations as a function of the retarded time. \emph{(Left panel):}
Comparison of the results after accounting for displacement-induced mode mixing to those obtained using 
half the extracted signal, both plain (i.e. the pure extracted signal), and with only the \textit{f}-mode part for SLy$-e0.50^{00}-$R4. The energy
estimate is only valid in the times between the periastron bursts.
\emph{(Middle and right panels):} The energy estimate computed accounting for displacement-induced mode mixing for the nonspinning
SLy and MS1b cases that have multiple encounters before merger, considering the R2 resolutions for which we have
data for the most eccentricities. In order to clarify these plots, we have excluded the portions of
the traces that are close to the periastron bursts.}
\label{fig:energy_estimate_illustration}
\end{figure*}

We compute the energy radiated by using the intrinsic $(2,\pm 2)$
modes of an individual star computed in the previous subsubsection.
Since the $f$-mode damping times are much longer
than the time between periastra, we assume that the $f$-mode GW signal is exactly 
sinusoidal, with angular frequency $\omega_{f\text{-mode}}$ and has the same amplitude in
both the $(2,2)$ and $(2,-2)$ modes, by symmetry. We then use this to compute
the antiderivative of $\Psi_4$ (by dividing by $\omega_{f\text{-mode}}$), which gives 
\begin{equation}
\dot{E} \simeq \frac{1}{8\pi} \left(\frac{r|\Psi^{4, f\text{-mode}}_{2,2}|}{\omega_{f\text{-mode}}}\right)^2,
\end{equation}
cf.\ Eq.~(52) in \cite{Brugmann:2008zz}.
%Here we have noted that we get equal contributions from the $(2, 2)$ and $(2, -2)$ modes.
Now, the energy stored in the $f$-mode oscillation
is $E_{f\text{-mode}}(t) = E_{f\text{-mode}}^\text{total} e^{-2t/\tau_{f\text{-mode}}}$.
The factor of 2 arises because we are looking at the energy, 
which goes as the amplitude squared. 
Thus, we have a radiated energy of $\dot{E}_\text{GW} = - \dot{E}_{f\text{-mode}} \simeq (2/\tau_{f\text{-mode}})E_{f\text{-mode}}^\text{total}$,
where we have evaluated this at $t = 0$, by the same argument
about the length of the \textit{f}-mode damping times compared to
the time between periastra as above. This finally gives
\begin{equation}
E_{f\text{-mode}}^\text{total} \simeq \frac{\tau_{f\text{-mode}}}{16\pi} \left( \frac{r|\Psi^{4, f\text{-mode}}_{2,2}|}{\omega_{f\text{-mode}}} \right)^2.
\end{equation}

We plot the energy estimate as a function of time for the nonspinning 
cases in Fig.~\ref{fig:energy_estimate_illustration}. 
(We do not include the spinning cases, since we found that 
the mode mixing removal was not working quite as well for them,
likely because the estimate of the \textit{f}-mode frequency is not sufficiently accurate.)
We illustrate how removing the displacement-induced mode mixing is 
necessary to make an accurate energy estimate, and then compare 
the energy estimates for the different cases.

We find that the amount of energy stored in the \textit{f}-mode 
oscillations increases with initial eccentricity, and is also larger for the MS1b stars than the
SLy stars, at a fixed eccentricity. 
Both of these are to be expected, since the periaston separations 
decrease with increasing eccentricity, so the stars are experiencing 
larger tidal perturbations, and the MS1b stars are
more tidally deformable than the SLy stars, so they will absorb more energy. 
We also find that the energy stored in the stars does not always 
increase monotonically with time, as would be expected, 
since the tidal perturbations of the stars may be close to out 
of phase with the already existing oscillations. This is also seen in the analytic calculations
in Ref.~\cite{Yang:2018bzx}. (There is a particularly dramatic
illustration of such an effect in Fig.~2 of Ref.~\cite{Radice:2016dwd}, though
that paper suggests that
it may be an artifact of the $\pi$ symmetry imposed during the evolution.)
The remaining time variation of the energy estimate in between 
periastron encounters is presumably due to the same effects 
discussed previously for the amplitude of the $(2,0)$ mode 
from displacement-induced mode mixing and the instantaneous frequency of the \textit{f}-mode signal.
The energy estimates for the first few encounters are robust across resolutions, while
the later ones differ more, since the later dynamics also differ between resolutions, as
shown for the gravitational waveform in Appendix~\ref{app:conv}.

We find that the \textit{f}-mode oscillations of the SLy stars store up to $\sim 10^{-3} M_\odot$
of energy in the $e = 0.5$ case. While $10^{-3} M_{\odot}$ is relatively small compared to some of the other energy scales in the problem
(such as the binding energy of the star or the binary's initial orbital binding energy,
which are on the order of $10^{-1}$ and $10^{-2} M_\odot$, respectively),
it is a tremendous amount of energy, $\sim 10^{51}$~erg, comparable
to the energy released in a supernova.

Furthermore, such energies will be sufficient to shatter the NSs'
crust and release the elastic energy stored in these oscillations
(cf.~\cite{Thompson:1995gw} where they report $\sim 10^{46}$ erg to be
stored in elastic energy). This will likely lead to flaring activity from
milliseconds up to possibly a few seconds before merger (cf.~Fig.~\ref{fig:energy_estimate_illustration}).
The signature could be similar to the resonance-induced cracking for quasi-circular
inspirals proposed in~\cite{Tsang:2011ad}, though through different mechanism and
time scales. Such a cracking of the NS crust is reported as one possible explanation for
sGRB precursors observed by Swift~\cite{Troja:2010zm} and might also be visible
for BNSs on eccentric orbits.

\subsection{Postmerger}

We analyze the GW spectrum of the postmerger waveform by performing a 
Fourier transform of the simulation data as discussed in Sec.~\ref{sec:simu}.

In Table~\ref{tab:postmerger}, we report the main peaks identified
in the postmerger PSDs from all our configurations including
results from different resolution simulations. We analyze 
the multipolar $r\Psi_4 ^{\lm}$ modes of the curvature scalar
and observe that modes other than (2,2) are also excited during the
postmerger phase.
These mode frequencies are labeled $f_1$, $f_2$, $f_3$
and are clearly harmonic, i.e.,~$f_1 \simeq f_2/2 \simeq f_3/3$;
cf.\ Ref.~\cite{Dietrich:2015pxa}.
These frequencies are extracted in the postmerger phase,
i.e., after the peak of the amplitude of the $(2,2)$ mode. For a clear interpretation,
we extract the $f_1$ frequency from the $(2, 1)$ mode and the $f_3$
frequency from the $(3, 3)$ mode, but they are present in
all the modes. The spectra are mainly characterized by
a dominant emission frequency $f_2$, related to the $(2,2)$ mode.

We also report the GW frequency at merger as $f_{\rm mrg}$. We find
that the dimensionless frequency at merger $M\omega_{\rm mrg}$ depends
on the EOS. While stiffer EOSs merge with a lower frequency,
softer EOSs merge at higher frequencies.
Furthermore, we observe a growing $m=1$ mode after the merger, as has been
found previously in both quasicircular and eccentric configurations, e.g., 
Refs.~\cite{Paschalidis:2015mla,Lehner:2016wjg,Radice:2016gym}. We find
that the $m = 1$ mode is an order-of-magnitude stronger in the SLy cases than in
the MS1b cases, but that $m = 1$ mode growth in the spinning cases is similar to
that in the irrotational cases.

In cases where a HMNS is formed
and in particular for configurations undergoing gravitational collapse
within dynamical times, the postmerger signal is shorter and the peaks
at specific frequencies $f_1$, $f_2$, $f_3$ are more difficult to extract
than for configurations that form MNSs. Overall, no direct correlation
between initial eccentricity and postmeger frequencies is observed.

\begin{table}[t]
\caption{
Post-merger properties. The columns give the name of the configuration, 
the dimensionless merger frequency $M\omega_{\rm mrg}$, the dimensionful merger frequency
$f_{\rm mrg}$ (in kHz), and the dominant postmerger frequencies extracted from the 
$(2,1)$, $(2,2)$, and $(3,3)$ modes. We mark ``$-$''
for cases where the frequencies could not be extracted properly, 
mostly due to the shorter lifetime of the HMNS.}
\label{tab:post-merger} 
\centering
\begin{tabular}{cccccc}
\toprule
Name & $M\omega_\text{mrg}$ & $f_{\rm mrg}$ & $f_1$ & $f_2$ & $f_3$ \\ 
     &                      & [kHz]          &  [kHz] &  [kHz] & [kHz] \\
\hline 
SLy$-e0.40^{00}-$R2 & $0.153$ & $1.83$ & $-$ & $3.51$ & $-$ \\
\hline
SLy$-e0.45^{00}-$R1 & $0.149$ & $1.78$ & $-$ & $3.45$ & $-$ \\

SLy$-e0.45^{00}-$R2 & $0.165$ & $1.97$ & $-$ & $3.64$ & $-$ \\

SLy$-e0.45^{00}-$R3 & $0.156$ & $1.87$ & $1.77$ & $3.54$ & $5.33$ \\

SLy$-e0.45^{00}-$R4 & $0.165$ & $1.97$ & $-$ & $3.66$ & $-$ \\
\hline
SLy$-e0.50^{00}-$R1 & $0.158$ & $1.89$ & $1.78$ & $3.56$ & $-$ \\

SLy$-e0.50^{00}-$R2 & $0.167$ & $1.99$ & $1.86$ & $3.69$ & $5.53$ \\

SLy$-e0.50^{00}-$R3 & $0.160$ & $1.91$ & $-$ & $-$ & $-$ \\

SLy$-e0.50^{00}-$R4 & $0.158$ & $1.89$ & $-$ & $3.47$ & $-$ \\
\hline
SLy$-e0.60^{00}-$R2 & $0.151$ & $1.81$ & $1.78$ & $3.56$ & $5.49$ \\
\toprule
SLy$-e0.40^{\uparrow \uparrow}-$R2 & $0.152$ & $1.82$ & $1.78$ & $3.54$ & $-$ \\
\hline
SLy$-e0.45^{\uparrow \uparrow}-$R1 & $0.140$ & $1.67$ & $1.81$ & $3.45$ & $-$ \\

SLy$-e0.45^{\uparrow \uparrow}-$R2 & $0.169$ & $2.01$ & $-$ & $3.46$ & $5.17$ \\

SLy$-e0.45^{\uparrow \uparrow}-$R3 & $0.177$ & $2.12$ & $-$ & $3.52$ & $-$ \\

SLy$-e0.45^{\uparrow \uparrow}-$R4 & $0.178$ & $2.12$ & $1.80$ & $3.62$ & $5.43$ \\
\hline
SLy$-e0.50^{\uparrow \uparrow}-$R2 & $0.157$ & $1.87$ & $1.87$ & $3.72$ & $5.62$ \\
\hline
SLy$-e0.60^{\uparrow \uparrow}-$R2 & $0.139$ & $1.65$ & $1.72$ & $3.47$ & $5.17$ \\
\toprule
MS1b$-e0.45^{00}-$R2 & $0.116$ & $1.36$ & $1.07$ & $2.06$ & $3.16$ \\
\hline
MS1b$-e0.50^{00}-$R2 & $0.112$ & $1.32$ & $1.05$ & $2.11$ & $3.16$ \\
\hline
MS1b$-e0.60^{00}-$R2 & $0.112$ & $1.32$ & $1.05$ & $2.11$ & $3.16$ \\
\toprule
MS1b$-e0.45^{\uparrow \uparrow}-$R2 & $0.128$ & $1.50$ & $1.08$ & $2.15$ & $3.21$ \\
\hline
MS1b$-e0.50^{\uparrow \uparrow}-$R2 & $0.126$ & $1.48$ & $1.07$ & $2.17$ & $3.12$ \\
\hline
MS1b$-e0.60^{\uparrow \uparrow}-$R2 & $0.106$ & $1.25$ & $1.10$ & $2.17$ & $3.24$ \\
\hline
\end{tabular}
\end{table}

\section{Summary}
\label{sec:summary}
In this article, we present and analyze a number of full GR numerical
simulations of eccentric BNS mergers with consistent ID employing 
either irrotational or aligned-spin stars.
We systematically vary the initial eccentricity in our
simulations to isolate the effect of (large) eccentricity with a fixed initial
separation of the NSs. Out of the total of 23 simulations
(including different physical configurations as well as different resolutions)
presented in this article, 21 of them have been made freely available in~\cite{Dietrich:2018phi,CoRe};
the remaining $2$ simulations will be made public in the near future. 
In the following we summarize our findings.\\

\paragraph*{\textbf{Dynamics:-}}
We find that depending on the initial eccentricity, 
the number of orbits significantly varies (starting from a fixed coordinate
separation of the stars), ranging from half an orbit for the most eccentric system 
to as many as $\sim 18$ orbits for the $e = 0.40$ configuration
employing the SLy EOS and aligned spins. Since some of the
simulations are evolved for more than 140~ms to capture the 
full dynamics of the system, 
these are among the longest full GR numerical evolutions of BNSs
performed to date (in particular {SLy$-e0.40^{\uparrow \uparrow}-$R2} with a length of $\approx 172\ {\rm ms}$);
see also~\cite{Haas:2016cop} and~\cite{DePietri:2018tpx} for the longest simulations 
of quasi-circular BNS inspirals, concentrating on the inspiral and postmerger phases, respectively.
For the configurations with aligned-spins, and in particular for systems
which undergo multiple non-merging encounters before the merger, we also find that more
angular momentum and energy is emitted before the merger,
compared to equivalent nonspinning configurations.
For the masses and EOSs we consider here, the merger remnant either
forms a stable MNS remnant or forms a 
HMNS which will eventually collapse to a BH. 
In fact, several evolved systems with SLy EOS form a BH even during the simulation time. 

As expected, the properties of the merger remnant are not only dependent 
on the physical properties such as EOS or initial intrinsic spin, but also
depend on the grid resolutions used to evolve the system.
Overall, we find that the measurement of the remnant's
lifetime is less robust for the eccentric simulations we
consider than for quasicircular orbits. One reason could be
the sensitive dependence of the postmerger evolution
on the number of close encounters before merger,
which itself depends on the eccentricity, spin,
and/or resolution. Furthermore, we do not find any clear imprint of
eccentricity on the merger
remnant properties in general. Thus,
quantitative statements must await future work 
when much higher resolution
evolutions are available. \\

\paragraph*{\textbf{Ejecta and EM counterparts:-}}
We successfully tested a new routine in BAM for computing unbound matter that
minimizes errors introduced in estimating ejecta mass due to
the presence of an artificial atmosphere.
Good agreement between the new method
and the old one with differences in the estimates of the unbound masses
below 11\% is achieved.

Even though we do not obtain clean convergence for the estimates of the
unbound matter with increasing resolution, our results are in good agreement with
the few comparable results available in the literature
~\cite{Stephens:2011as,East:2012ww,East:2015vix,East:2016zvv,Radice:2016dwd}.
Specifically, $\sim\mathcal{O}(10^{-2})~M_{\odot}$ of matter can be ejected at the merger.
This is slightly more than in the quasicircular case for the SLy binaries, and about an order of magnitude more
for the MS1b binaries. In our simulations tidal tail ejecta are more prominent compared 
to shock-heated ejecta or ejecta due to the redistribution
of angular momentum in the postmerger remnant. Moreover, unbound
matter $\sim\mathcal{O}(10^{-3})-\mathcal{O}(10^{-2})~M_{\odot}$ is ejected as
a mildly relativistic and mildly isotropic outflow with velocities $\sim 6-15$\%
of the speed of light.

For EM transients we find compatible results for quasicircular and eccentric
BNS mergers. In general, the considered configurations will produce kilonovae 
with luminosities between $10^{39}-10^{42}$ erg~s$^{-1}$ over a time ranging from a few days
to two weeks after the merger. On the other hand, the radio flares will have
the largest fluence at $t^\text{radio} _\text{peak}$ $\sim\mathcal{O}(\text{years})$,
similar to equivalent quasicircular cases.

Moreover, in contrast to noneccentric mergers, we find that unbound matter of
$\sim \mathcal{O}(10^{-4})-\mathcal{O}(10^{-3})~M_{\odot}$ of neutron rich material can be ejected
before the merger i.e. during the binary's successive periastron encounters. This would in principle allow for
observations of EM emissions before the merger, although observatories would
require early notice.\\

\paragraph*{\textbf{Gravitational Waves:-}}
A notable
feature for BNSs on eccentric orbits is the superposition
of gravitational waves from the quasinormal modes of the NSs (specifically the \textit{f}-mode) on the GW signal from the binary's
orbital motion. These quasinormal modes are excited by the time-varying tidal perturbations of the stars during their periastron passages.

We find good agreement between the
\textit{f}-mode frequency from our simulations  for the irrotational cases
and the one obtained from the
perturbation theory estimate for an equivalent isolated NS.
The $(2,0)$ \textit{f}-mode signal found in our simulations is accounted for
entirely by mode mixing of the intrinsic $(2, \pm 2)$ modes of the stars due to
the stars' displacement from the origin. 
We also estimate
the energy stored in the \textit{f}-mode oscillations and find that it increases
with increasing eccentricity. In general, stiff EOSs, as MS1b, store more energy in the
oscillations compared to soft EOSs, e.g.~SLy. Additionally, the energy stored
in the oscillations of the stars does not always increase monotonically with time.
This is to be expected, since for some encounters the tidal perturbations will be
out of phase with the already existing oscillations. Overall, these
oscillations can store $\sim \mathcal{O}(10^{-8})-\mathcal{O}(10^{-3})~M_{\odot}$
of energy depending on eccentricity and the sequence of non-merging encounters.

We find the same qualitative relation between the merger frequency and
the stiffness of the EOS that is known for quasicircular binaries, where binaries constructed using
the stiffer MS1b EOS merge at lower frequencies than those constructed using the softer SLy EOS.
In the postmerger signal modes other than $(2,\pm 2)$ modes are also excited
and the same harmonic relation between the dominant frequencies found for
quasicircular binaries holds.

With regard to the observability of eccentric BNS mergers,
the most prominent features are the burst of gravitational
radiation associated with each close encounter, which might be observable 
with future 3G detectors. On the other hand, observing
the \textit{f}-mode oscillations might require even higher sensitivities 
or fortuitous circumstances for 3G detectors. 
If observable, an interesting and notable characteristic would be the 
change in the \textit{f}-mode amplitude after each encounter, 
which may increase or decrease as discussed in Sec.~\ref{sec:GWs}.

\begin{acknowledgments}
  We thank Roland Haas and the anonymous referee for a careful reading and helpful comments on the manuscript.
  It is also a pleasure to thank S.~Bernuzzi, R.~Dudi, R.~Gold, T.~Hinderer, and J.~Steinhoff 
  for useful comments and discussion.
  S.~V.~C.~was supported by the DFG Research Training 
  Group 1523/2 ``Quantum and Gravitational Fields.''
  T.~D.\ acknowledges support by the European Union's Horizon 
  2020 research and innovation program under grant
  agreement No 749145, BNSmergers.
  N.~K.~J.-M.\ acknowledges
  support from STFC Consolidator Grant No.~ST/L000636/1.
  B.~B.\ was supported by DFG Grant No. BR 2176/5-1.
  W.~T.\ was supported by the National Science Foundation
  under grant PHY-1707227.
  Also, this work has received
  funding from the European Union's Horizon 2020 research and innovation programme under
  the Marie Sklodowska-Curie Grant Agreement No.~690904. This research was supported in
  part by Perimeter Institute for Theoretical Physics. Research at Perimeter Institute is supported by
  the Government of Canada through Industry Canada and by the Province of Ontario through the
  Ministry of Economic Development \& Innovation.
  Computations were performed on 
  the supercomputer SuperMUC at the LRZ
  (Munich) under the project number pr48pu
  and on the ARA cluster of the University of Jena.
\end{acknowledgments}

\appendix

\begin{figure}[t]
\includegraphics[width=0.5\textwidth]{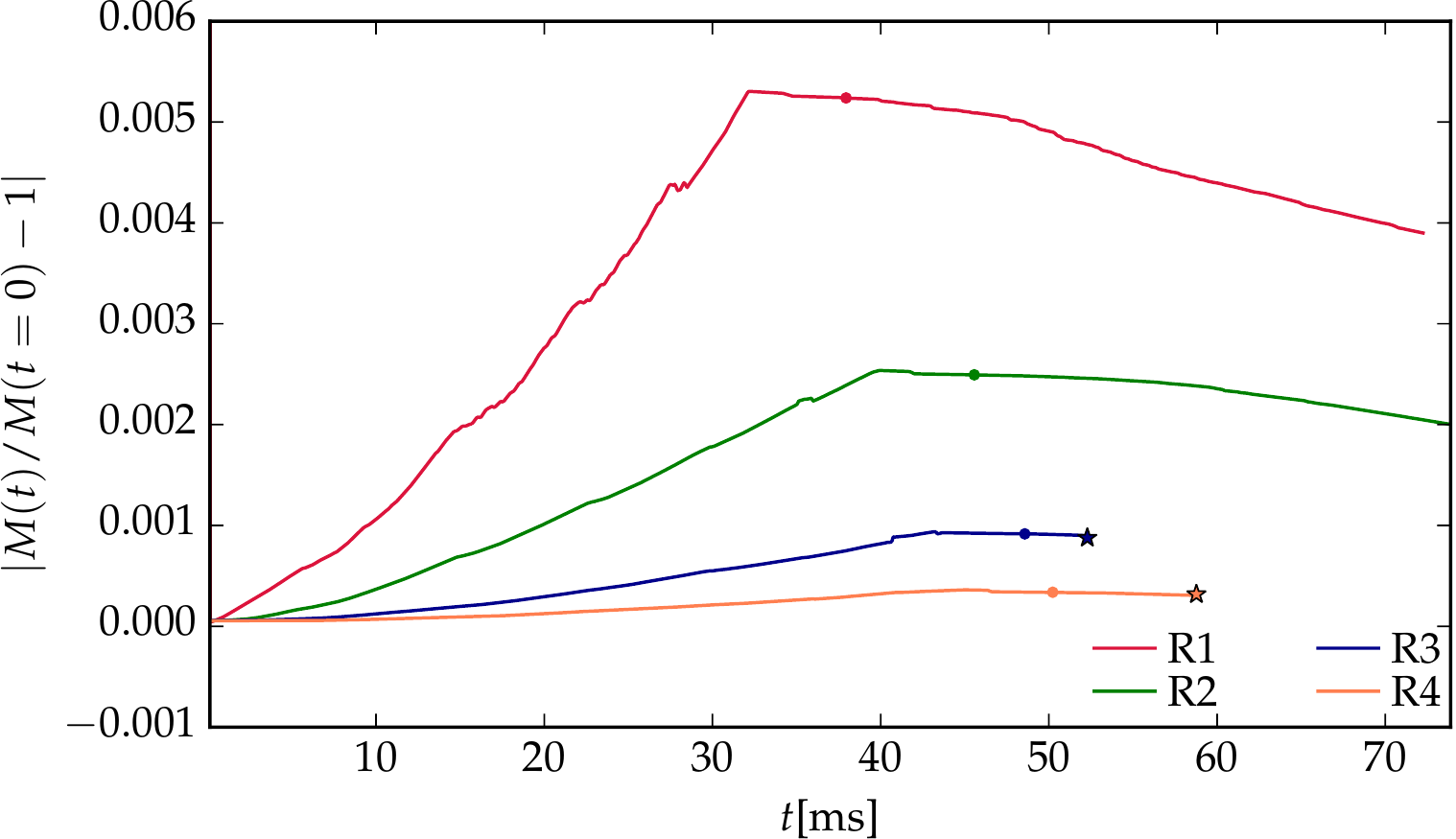}
\caption{Rest mass conservation of SLy$-e0.50^{00}$ case on refinement 
level $l=1$.
The panel shows the absolute error of the rescaled mass 
evolution of the baryonic mass. We mark the moment of merger as 
circles and the moment when a BH forms with a star.
The error in the total mass stays below 0.6\%
for the lowest resolution and below 0.01\% for the highest resolution
over the entire evolution time.
Mass conservation is not violated once a BH forms.}
\label{fig:convergence:mass}
\end{figure}

\begin{figure}[t]
\includegraphics[width=0.5\textwidth]{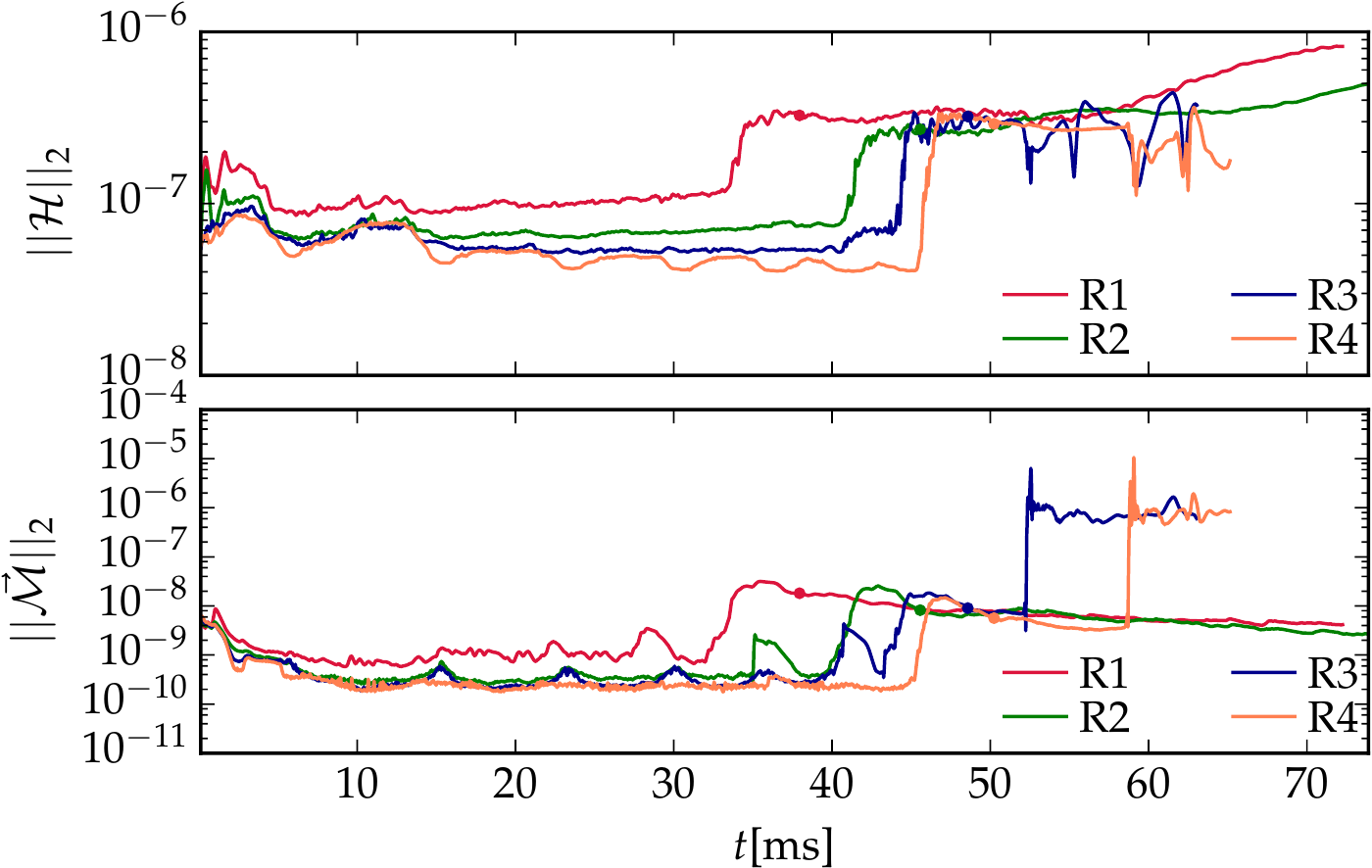}
\caption{ADM constraints for the SLy$-e0.50^{00}$ case. 
The upper panel shows the $L^{2}$ volume norm of the Hamiltonian 
constraint, $||\mathcal{H}||_2$. The lower panel shows the
Euclidean norm of the $L^{2}$ volume norms of the Cartesian components of the
momentum constraint, $||\vec{\mathcal{M}}||_2 = \sqrt{||\mathcal{M}^x||^2 _2 +
||\mathcal{M}^y||^2 _2 + ||\mathcal{M}^z||^2 _2}$.
The constraints are evaluated on refinement level 4 and 
are decreasing for increasing resolution 
during the inspiral of the neutron stars.
Note that here we restrict our convergence analysis
to the refinement level 4, cf. Tab.~\ref{tab:config}.
This is needed since for the coarser refinement levels even
the higher resolution setups are only barely able to resolve the star
whereas for the coarser resolutions not a single
point covers the NS leading
to a nonconvergent behavior.
We have filtered 
our data with an average filter to give a better 
visualization and reduce high frequency noise and mark the times of merger by circles.}
\label{fig:convergence:hammomlev4}
\end{figure}

\begin{figure}[t]
\includegraphics[width=0.5\textwidth]{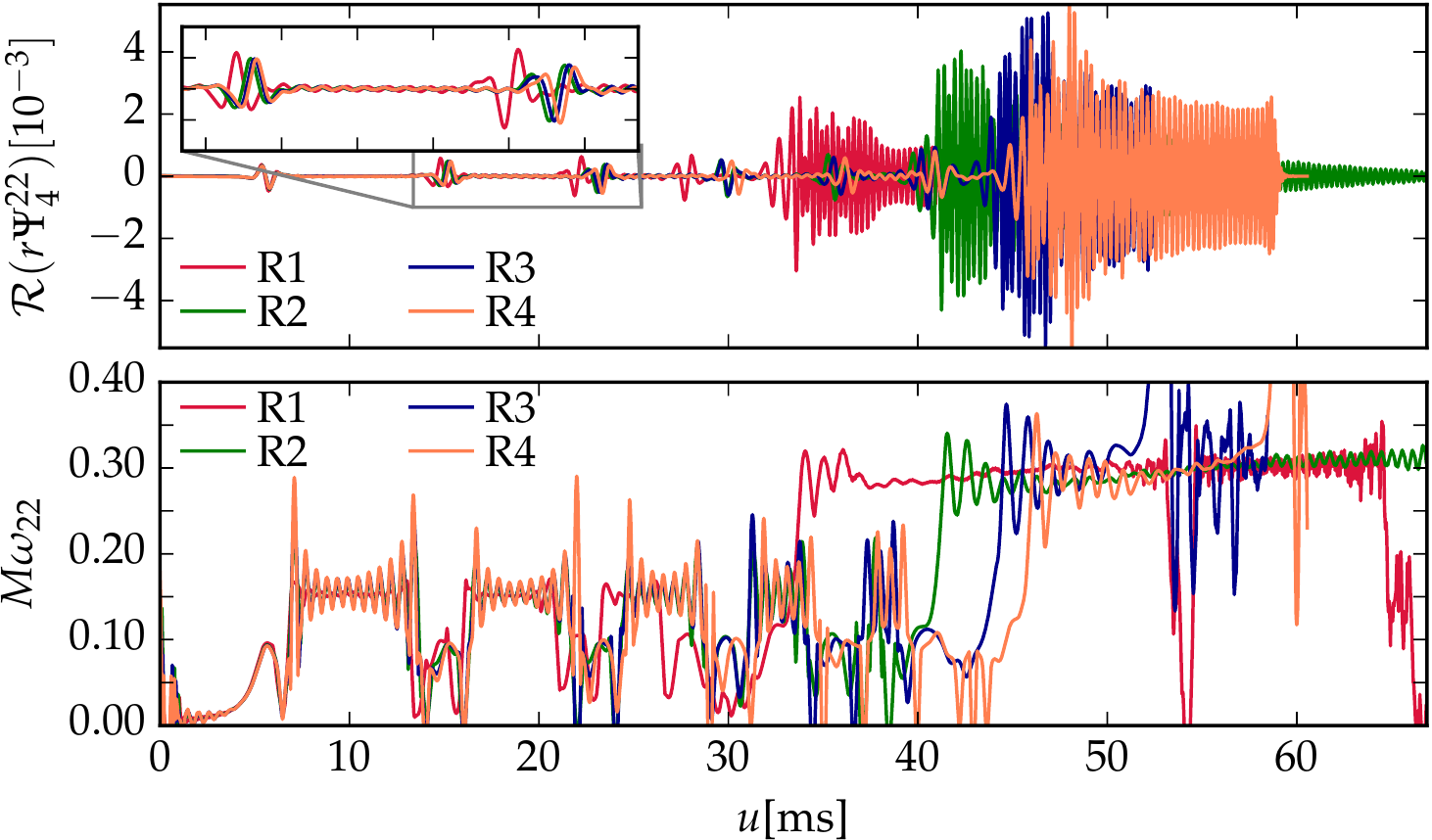}
\caption{(\textit{Top panel}): Real part of the curvature multipole
$r\Psi ^{22} _4$
for the SLy$-e0.50^{00}$ configuration for four different resolution,
see Table~\ref{tab:grid}.
The inset figure shows two periastron encounters and the region in between
where the \textit{f}-mode oscillations are present.
(\textit{Bottom panel}): The instantaneous dimensionless
frequency $M\omega _{22}$ computed from $r\Psi ^{22} _4$.
The plots show the robustness of the NS oscillations and the different evolution
of the postmerger part. The merger times also vary depending on the resolution due
to varying numerical dissipation, which is an inevitable numerical phenomenon for all
simulations.}
\label{fig:convergence:waveform}
\end{figure}

\section{Convergence Study}
\label{app:conv}
To give some diagnostics for the 
accuracy of our simulations, we present 
a convergence study for the conservation of 
baryonic mass of the system,
Fig.~\ref{fig:convergence:mass}, the
{\rm ADM} constraints,
Fig.~\ref{fig:convergence:hammomlev4}, and the waveform,
Fig.~\ref{fig:convergence:waveform}, for the SLy$-e0.50^{00}$ 
configuration. 
We refer the reader to Refs.~\cite{Dietrich:2015pxa,Tichy:2009yr} 
for a detailed discussion about the convergence and accuracy 
of SGRID and Refs.~\cite{Bernuzzi:2011aq,Dietrich:2015iva,Bernuzzi:2016pie} 
for the accuracy of BAM.\\

\paragraph*{\textbf{Mass conservation:-}} 
A detailed discussion about the mass conservation in BNS simulations with BAM 
was presented in Ref.~\cite{Dietrich:2015iva}.
Here, we want to present the rest mass
conservation for at least one of
our simulations. Figure~\ref{fig:convergence:mass}
shows the mass conservation for the SLy$-e0.50^{00}$
configuration. It is clear that the mass
conservation improves with increasing resolution.
However, note that the merger itself happens at
different times due to different numerical dissipation
for different resolutions. Additionally, after the merger no considerable
mass loss is present at higher resolutions (except for the
cases where a BH forms, which is expected) and
the difference to the initial mass stays below 0.6\%
for the lowest resolution simulation
and below 0.01\% for the highest resolution simulation.\\

\paragraph*{\textbf{{ADM} constraints:-}} 
Since we use a free evolution scheme of the (3+1)-decomposed 
Einstein equations, we have to ensure that
the Hamiltonian and the momentum constraints are fulfilled
over the entire simulation. While we explicitly solve
the constraints to obtain our initial data, the
constraints are not solved during the simulation.
Figure~\ref{fig:convergence:hammomlev4} shows the L$^2$ 
volume norm of the Hamiltonian constraint (top panel)
and the L$^2$ volume norm of the square magnitude
of the momentum constraint (bottom panel) during the
simulation. We see that the constraints
are well behaved for the entire duration of the
numerical simulation. Oscillations during
the inspiral are caused by the movement of
inner refinement levels which follow the motion of 
the neutron stars. After merger those oscillations are 
absent since the stars stay near the center or
only move with a small velocity compared to
the inspiral.\\

\paragraph*{\textbf{Waveform:-}}
In order to estimate the convergence properties of the GW signal, we
use the merger times for the SLy$-e0.50^{00}$ case with
multiple resolutions (cf.\ Fig.~\ref{fig:convergence:waveform}) 
presented in this paper.

Considering the merger times computed from
the R1, R2, R3, and R4 resolutions, we obtain a convergence
of order $\sim 1.5$. While this is an indicator that the simulations are 
in the convergent regime a more careful investigations, see e.g.~\cite{Bernuzzi:2016pie}, 
is needed for a full error assessment. 
Nevertheless, our results show that the simulations
can be used as first test beds once eccentric BNS waveform models
become available. But in order to perform detailed waveform
modelling and further reduce the uncertainty of
our numerical results, even higher resolution simulations
are the immediate need. These high resolution
simulations will require large amounts of computational
resources but are currently underway.

%%______________________________________________________________

\bibliography{paper20181112.bbl}

\end{document}